\documentclass[12pt,prd,showpacs,tightenlines,nofootinbib]{revtex4}
\usepackage{bm}
\usepackage{graphics}
\usepackage{rotating}
\usepackage{epsfig}

\begin{document}
\begin{flushright}
{\bf Preprint SSU-HEP-04/12\\
Samara State University}
\end{flushright}

\title{THEORY OF MUONIC HYDROGEN - MUONIC\\
DEUTERIUM ISOTOPE SHIFT}

\author{\firstname{A.P.} \surname{Martynenko}}
\email{mart@ssu.samara.ru(A.P.Martynenko)}
\affiliation{Samara State University,\\
443011, Ac. Pavlov 1, Samara, Russia}

\begin{abstract}
We calculate the corrections of orders $\alpha^3$, $\alpha^4$ and $\alpha^5$
to the Lamb shift of the 1S and 2S energy levels of muonic hydrogen $(\mu p)$
and muonic deuterium $(\mu d)$. The nuclear structure effects are taken into
account in terms of the proton $r_p$ and deuteron $r_d$ charge radii for the
one-photon interaction and by means of the proton and deuteron
electromagnetic form factors in the case of one-loop amplitudes. The obtained
numerical value of the isotope shift $(\mu d)$ - $(\mu p)$ for the splitting
$(1S\div 2S)$ 101003.3495 meV can be considered as a reliable estimation for
corresponding experiment with the accuracy $10^{-6}$. The fine structure
interval $E(1S)-8E(2S)$ in muonic hydrogen and muonic deuterium are
calculated.
\end{abstract}


\maketitle

\immediate\write16{<<WARNING: LINEDRAW macros work with emTeX-dvivers
                    and other drivers supporting emTeX \special's
                    (dviscr, dvihplj, dvidot, dvips, dviwin, etc.) >>}

\newdimen\Lengthunit       \Lengthunit  = 1.5cm
\newcount\Nhalfperiods     \Nhalfperiods= 9
\newcount\magnitude        \magnitude = 1000

\catcode`\*=11
\newdimen\L*   \newdimen\d*   \newdimen\d**
\newdimen\dm*  \newdimen\dd*  \newdimen\dt*
\newdimen\a*   \newdimen\b*   \newdimen\c*
\newdimen\a**  \newdimen\b**
\newdimen\xL*  \newdimen\yL*
\newdimen\rx*  \newdimen\ry*
\newdimen\tmp* \newdimen\linwid*

\newcount\k*   \newcount\l*   \newcount\m*
\newcount\k**  \newcount\l**  \newcount\m**
\newcount\n*   \newcount\dn*  \newcount\r*
\newcount\N*   \newcount\*one \newcount\*two  \*one=1 \*two=2
\newcount\*ths \*ths=1000
\newcount\angle*  \newcount\q*  \newcount\q**
\newcount\angle** \angle**=0
\newcount\sc*     \sc*=0

\newtoks\cos*  \cos*={1}
\newtoks\sin*  \sin*={0}

\catcode`\[=13

\def\rotate(#1){\advance\angle**#1\angle*=\angle**
\q**=\angle*\ifnum\q**<0\q**=-\q**\fi
\ifnum\q**>360\q*=\angle*\divide\q*360\multiply\q*360\advance\angle*-\q*\fi
\ifnum\angle*<0\advance\angle*360\fi\q**=\angle*\divide\q**90\q**=\q**
\def\sgcos*{+}\def\sgsin*{+}\relax
\ifcase\q**\or
 \def\sgcos*{-}\def\sgsin*{+}\or
 \def\sgcos*{-}\def\sgsin*{-}\or
 \def\sgcos*{+}\def\sgsin*{-}\else\fi
\q*=\q**
\multiply\q*90\advance\angle*-\q*
\ifnum\angle*>45\sc*=1\angle*=-\angle*\advance\angle*90\else\sc*=0\fi
\def[##1,##2]{\ifnum\sc*=0\relax
\edef\cs*{\sgcos*.##1}\edef\sn*{\sgsin*.##2}\ifcase\q**\or
 \edef\cs*{\sgcos*.##2}\edef\sn*{\sgsin*.##1}\or
 \edef\cs*{\sgcos*.##1}\edef\sn*{\sgsin*.##2}\or
 \edef\cs*{\sgcos*.##2}\edef\sn*{\sgsin*.##1}\else\fi\else
\edef\cs*{\sgcos*.##2}\edef\sn*{\sgsin*.##1}\ifcase\q**\or
 \edef\cs*{\sgcos*.##1}\edef\sn*{\sgsin*.##2}\or
 \edef\cs*{\sgcos*.##2}\edef\sn*{\sgsin*.##1}\or
 \edef\cs*{\sgcos*.##1}\edef\sn*{\sgsin*.##2}\else\fi\fi
\cos*={\cs*}\sin*={\sn*}\global\edef\gcos*{\cs*}\global\edef\gsin*{\sn*}}\relax
\ifcase\angle*[9999,0]\or
[999,017]\or[999,034]\or[998,052]\or[997,069]\or[996,087]\or
[994,104]\or[992,121]\or[990,139]\or[987,156]\or[984,173]\or
[981,190]\or[978,207]\or[974,224]\or[970,241]\or[965,258]\or
[961,275]\or[956,292]\or[951,309]\or[945,325]\or[939,342]\or
[933,358]\or[927,374]\or[920,390]\or[913,406]\or[906,422]\or
[898,438]\or[891,453]\or[882,469]\or[874,484]\or[866,499]\or
[857,515]\or[848,529]\or[838,544]\or[829,559]\or[819,573]\or
[809,587]\or[798,601]\or[788,615]\or[777,629]\or[766,642]\or
[754,656]\or[743,669]\or[731,681]\or[719,694]\or[707,707]\or
\else[9999,0]\fi}

\catcode`\[=12

\def\GRAPH(hsize=#1)#2{\hbox to #1\Lengthunit{#2\hss}}

\def\Linewidth#1{\global\linwid*=#1\relax
\global\divide\linwid*10\global\multiply\linwid*\mag
\global\divide\linwid*100\special{em:linewidth \the\linwid*}}

\Linewidth{.4pt}
\def\sm*{\special{em:moveto}}
\def\sl*{\special{em:lineto}}
\let\moveto=\sm*
\let\lineto=\sl*
\newbox\spm*   \newbox\spl*
\setbox\spm*\hbox{\sm*}
\setbox\spl*\hbox{\sl*}

\def\mov#1(#2,#3)#4{\rlap{\L*=#1\Lengthunit
\xL*=#2\L* \yL*=#3\L*
\xL*=\xscale\xL* \yL*=\yscale\yL*
\rx* \the\cos*\xL* \tmp* \the\sin*\yL* \advance\rx*-\tmp*
\ry* \the\cos*\yL* \tmp* \the\sin*\xL* \advance\ry*\tmp*
\kern\rx*\raise\ry*\hbox{#4}}}

\def\rmov*(#1,#2)#3{\rlap{\xL*=#1\yL*=#2\relax
\rx* \the\cos*\xL* \tmp* \the\sin*\yL* \advance\rx*-\tmp*
\ry* \the\cos*\yL* \tmp* \the\sin*\xL* \advance\ry*\tmp*
\kern\rx*\raise\ry*\hbox{#3}}}

\def\lin#1(#2,#3){\rlap{\sm*\mov#1(#2,#3){\sl*}}}

\def\arr*(#1,#2,#3){\rmov*(#1\dd*,#1\dt*){\sm*
\rmov*(#2\dd*,#2\dt*){\rmov*(#3\dt*,-#3\dd*){\sl*}}\sm*
\rmov*(#2\dd*,#2\dt*){\rmov*(-#3\dt*,#3\dd*){\sl*}}}}

\def\arrow#1(#2,#3){\rlap{\lin#1(#2,#3)\mov#1(#2,#3){\relax
\d**=-.012\Lengthunit\dd*=#2\d**\dt*=#3\d**
\arr*(1,10,4)\arr*(3,8,4)\arr*(4.8,4.2,3)}}}

\def\arrlin#1(#2,#3){\rlap{\L*=#1\Lengthunit\L*=.5\L*
\lin#1(#2,#3)\rmov*(#2\L*,#3\L*){\arrow.1(#2,#3)}}}

\def\dasharrow#1(#2,#3){\rlap{{\Lengthunit=0.9\Lengthunit
\dashlin#1(#2,#3)\mov#1(#2,#3){\sm*}}\mov#1(#2,#3){\sl*
\d**=-.012\Lengthunit\dd*=#2\d**\dt*=#3\d**
\arr*(1,10,4)\arr*(3,8,4)\arr*(4.8,4.2,3)}}}

\def\clap#1{\hbox to 0pt{\hss #1\hss}}

\def\ind(#1,#2)#3{\rlap{\L*=.1\Lengthunit
\xL*=#1\L* \yL*=#2\L*
\rx* \the\cos*\xL* \tmp* \the\sin*\yL* \advance\rx*-\tmp*
\ry* \the\cos*\yL* \tmp* \the\sin*\xL* \advance\ry*\tmp*
\kern\rx*\raise\ry*\hbox{\lower2pt\clap{$#3$}}}}

\def\sh*(#1,#2)#3{\rlap{\dm*=\the\n*\d**
\xL*=\xscale\dm* \yL*=\yscale\dm* \xL*=#1\xL* \yL*=#2\yL*
\rx* \the\cos*\xL* \tmp* \the\sin*\yL* \advance\rx*-\tmp*
\ry* \the\cos*\yL* \tmp* \the\sin*\xL* \advance\ry*\tmp*
\kern\rx*\raise\ry*\hbox{#3}}}

\def\calcnum*#1(#2,#3){\a*=1000sp\b*=1000sp\a*=#2\a*\b*=#3\b*
\ifdim\a*<0pt\a*-\a*\fi\ifdim\b*<0pt\b*-\b*\fi
\ifdim\a*>\b*\c*=.96\a*\advance\c*.4\b*
\else\c*=.96\b*\advance\c*.4\a*\fi
\k*\a*\multiply\k*\k*\l*\b*\multiply\l*\l*
\m*\k*\advance\m*\l*\n*\c*\r*\n*\multiply\n*\n*
\dn*\m*\advance\dn*-\n*\divide\dn*2\divide\dn*\r*
\advance\r*\dn*
\c*=\the\Nhalfperiods5sp\c*=#1\c*\ifdim\c*<0pt\c*-\c*\fi
\multiply\c*\r*\N*\c*\divide\N*10000}

\def\dashlin#1(#2,#3){\rlap{\calcnum*#1(#2,#3)\relax
\d**=#1\Lengthunit\ifdim\d**<0pt\d**-\d**\fi
\divide\N*2\multiply\N*2\advance\N*\*one
\divide\d**\N*\sm*\n*\*one\sh*(#2,#3){\sl*}\loop
\advance\n*\*one\sh*(#2,#3){\sm*}\advance\n*\*one
\sh*(#2,#3){\sl*}\ifnum\n*<\N*\repeat}}

\def\dashdotlin#1(#2,#3){\rlap{\calcnum*#1(#2,#3)\relax
\d**=#1\Lengthunit\ifdim\d**<0pt\d**-\d**\fi
\divide\N*2\multiply\N*2\advance\N*1\multiply\N*2\relax
\divide\d**\N*\sm*\n*\*two\sh*(#2,#3){\sl*}\loop
\advance\n*\*one\sh*(#2,#3){\kern-1.48pt\lower.5pt\hbox{\rm.}}\relax
\advance\n*\*one\sh*(#2,#3){\sm*}\advance\n*\*two
\sh*(#2,#3){\sl*}\ifnum\n*<\N*\repeat}}

\def\shl*(#1,#2)#3{\kern#1#3\lower#2#3\hbox{\unhcopy\spl*}}

\def\trianglin#1(#2,#3){\rlap{\toks0={#2}\toks1={#3}\calcnum*#1(#2,#3)\relax
\dd*=.57\Lengthunit\dd*=#1\dd*\divide\dd*\N*
\divide\dd*\*ths \multiply\dd*\magnitude
\d**=#1\Lengthunit\ifdim\d**<0pt\d**-\d**\fi
\multiply\N*2\divide\d**\N*\sm*\n*\*one\loop
\shl**{\dd*}\dd*-\dd*\advance\n*2\relax
\ifnum\n*<\N*\repeat\n*\N*\shl**{0pt}}}

\def\wavelin#1(#2,#3){\rlap{\toks0={#2}\toks1={#3}\calcnum*#1(#2,#3)\relax
\dd*=.23\Lengthunit\dd*=#1\dd*\divide\dd*\N*
\divide\dd*\*ths \multiply\dd*\magnitude
\d**=#1\Lengthunit\ifdim\d**<0pt\d**-\d**\fi
\multiply\N*4\divide\d**\N*\sm*\n*\*one\loop
\shl**{\dd*}\dt*=1.3\dd*\advance\n*\*one
\shl**{\dt*}\advance\n*\*one
\shl**{\dd*}\advance\n*\*two
\dd*-\dd*\ifnum\n*<\N*\repeat\n*\N*\shl**{0pt}}}

\def\w*lin(#1,#2){\rlap{\toks0={#1}\toks1={#2}\d**=\Lengthunit\dd*=-.12\d**
\divide\dd*\*ths \multiply\dd*\magnitude
\N*8\divide\d**\N*\sm*\n*\*one\loop
\shl**{\dd*}\dt*=1.3\dd*\advance\n*\*one
\shl**{\dt*}\advance\n*\*one
\shl**{\dd*}\advance\n*\*one
\shl**{0pt}\dd*-\dd*\advance\n*1\ifnum\n*<\N*\repeat}}

\def\l*arc(#1,#2)[#3][#4]{\rlap{\toks0={#1}\toks1={#2}\d**=\Lengthunit
\dd*=#3.037\d**\dd*=#4\dd*\dt*=#3.049\d**\dt*=#4\dt*\ifdim\d**>10mm\relax
\d**=.25\d**\n*\*one\shl**{-\dd*}\n*\*two\shl**{-\dt*}\n*3\relax
\shl**{-\dd*}\n*4\relax\shl**{0pt}\else
\ifdim\d**>5mm\d**=.5\d**\n*\*one\shl**{-\dt*}\n*\*two
\shl**{0pt}\else\n*\*one\shl**{0pt}\fi\fi}}

\def\d*arc(#1,#2)[#3][#4]{\rlap{\toks0={#1}\toks1={#2}\d**=\Lengthunit
\dd*=#3.037\d**\dd*=#4\dd*\d**=.25\d**\sm*\n*\*one\shl**{-\dd*}\relax
\n*3\relax\sh*(#1,#2){\xL*=\xscale\dd*\yL*=\yscale\dd*
\kern#2\xL*\lower#1\yL*\hbox{\sm*}}\n*4\relax\shl**{0pt}}}

\def\shl**#1{\c*=\the\n*\d**\d*=#1\relax
\a*=\the\toks0\c*\b*=\the\toks1\d*\advance\a*-\b*
\b*=\the\toks1\c*\d*=\the\toks0\d*\advance\b*\d*
\a*=\xscale\a*\b*=\yscale\b*
\rx* \the\cos*\a* \tmp* \the\sin*\b* \advance\rx*-\tmp*
\ry* \the\cos*\b* \tmp* \the\sin*\a* \advance\ry*\tmp*
\raise\ry*\rlap{\kern\rx*\unhcopy\spl*}}

\def\wlin*#1(#2,#3)[#4]{\rlap{\toks0={#2}\toks1={#3}\relax
\c*=#1\l*\c*\c*=.01\Lengthunit\m*\c*\divide\l*\m*
\c*=\the\Nhalfperiods5sp\multiply\c*\l*\N*\c*\divide\N*\*ths
\divide\N*2\multiply\N*2\advance\N*\*one
\dd*=.002\Lengthunit\dd*=#4\dd*\multiply\dd*\l*\divide\dd*\N*
\divide\dd*\*ths \multiply\dd*\magnitude
\d**=#1\multiply\N*4\divide\d**\N*\sm*\n*\*one\loop
\shl**{\dd*}\dt*=1.3\dd*\advance\n*\*one
\shl**{\dt*}\advance\n*\*one
\shl**{\dd*}\advance\n*\*two
\dd*-\dd*\ifnum\n*<\N*\repeat\n*\N*\shl**{0pt}}}

\def\wavebox#1{\setbox0\hbox{#1}\relax
\a*=\wd0\advance\a*14pt\b*=\ht0\advance\b*\dp0\advance\b*14pt\relax
\hbox{\kern9pt\relax
\rmov*(0pt,\ht0){\rmov*(-7pt,7pt){\wlin*\a*(1,0)[+]\wlin*\b*(0,-1)[-]}}\relax
\rmov*(\wd0,-\dp0){\rmov*(7pt,-7pt){\wlin*\a*(-1,0)[+]\wlin*\b*(0,1)[-]}}\relax
\box0\kern9pt}}

\def\rectangle#1(#2,#3){\relax
\lin#1(#2,0)\lin#1(0,#3)\mov#1(0,#3){\lin#1(#2,0)}\mov#1(#2,0){\lin#1(0,#3)}}

\def\dashrectangle#1(#2,#3){\dashlin#1(#2,0)\dashlin#1(0,#3)\relax
\mov#1(0,#3){\dashlin#1(#2,0)}\mov#1(#2,0){\dashlin#1(0,#3)}}

\def\waverectangle#1(#2,#3){\L*=#1\Lengthunit\a*=#2\L*\b*=#3\L*
\ifdim\a*<0pt\a*-\a*\def\x*{-1}\else\def\x*{1}\fi
\ifdim\b*<0pt\b*-\b*\def\y*{-1}\else\def\y*{1}\fi
\wlin*\a*(\x*,0)[-]\wlin*\b*(0,\y*)[+]\relax
\mov#1(0,#3){\wlin*\a*(\x*,0)[+]}\mov#1(#2,0){\wlin*\b*(0,\y*)[-]}}

\def\calcparab*{\ifnum\n*>\m*\k*\N*\advance\k*-\n*\else\k*\n*\fi
\a*=\the\k* sp\a*=10\a*\b*\dm*\advance\b*-\a*\k*\b*
\a*=\the\*ths\b*\divide\a*\l*\multiply\a*\k*
\divide\a*\l*\k*\*ths\r*\a*\advance\k*-\r*\dt*=\the\k*\L*}

\def\arcto#1(#2,#3)[#4]{\rlap{\toks0={#2}\toks1={#3}\calcnum*#1(#2,#3)\relax
\dm*=135sp\dm*=#1\dm*\d**=#1\Lengthunit\ifdim\dm*<0pt\dm*-\dm*\fi
\multiply\dm*\r*\a*=.3\dm*\a*=#4\a*\ifdim\a*<0pt\a*-\a*\fi
\advance\dm*\a*\N*\dm*\divide\N*10000\relax
\divide\N*2\multiply\N*2\advance\N*\*one
\L*=-.25\d**\L*=#4\L*\divide\d**\N*\divide\L*\*ths
\m*\N*\divide\m*2\dm*=\the\m*5sp\l*\dm*\sm*\n*\*one\loop
\calcparab*\shl**{-\dt*}\advance\n*1\ifnum\n*<\N*\repeat}}

\def\arrarcto#1(#2,#3)[#4]{\L*=#1\Lengthunit\L*=.54\L*
\arcto#1(#2,#3)[#4]\rmov*(#2\L*,#3\L*){\d*=.457\L*\d*=#4\d*\d**-\d*
\rmov*(#3\d**,#2\d*){\arrow.02(#2,#3)}}}

\def\dasharcto#1(#2,#3)[#4]{\rlap{\toks0={#2}\toks1={#3}\relax
\calcnum*#1(#2,#3)\dm*=\the\N*5sp\a*=.3\dm*\a*=#4\a*\ifdim\a*<0pt\a*-\a*\fi
\advance\dm*\a*\N*\dm*
\divide\N*20\multiply\N*2\advance\N*1\d**=#1\Lengthunit
\L*=-.25\d**\L*=#4\L*\divide\d**\N*\divide\L*\*ths
\m*\N*\divide\m*2\dm*=\the\m*5sp\l*\dm*
\sm*\n*\*one\loop\calcparab*
\shl**{-\dt*}\advance\n*1\ifnum\n*>\N*\else\calcparab*
\sh*(#2,#3){\xL*=#3\dt* \yL*=#2\dt*
\rx* \the\cos*\xL* \tmp* \the\sin*\yL* \advance\rx*\tmp*
\ry* \the\cos*\yL* \tmp* \the\sin*\xL* \advance\ry*-\tmp*
\kern\rx*\lower\ry*\hbox{\sm*}}\fi
\advance\n*1\ifnum\n*<\N*\repeat}}

\def\*shl*#1{\c*=\the\n*\d**\advance\c*#1\a**\d*\dt*\advance\d*#1\b**
\a*=\the\toks0\c*\b*=\the\toks1\d*\advance\a*-\b*
\b*=\the\toks1\c*\d*=\the\toks0\d*\advance\b*\d*
\rx* \the\cos*\a* \tmp* \the\sin*\b* \advance\rx*-\tmp*
\ry* \the\cos*\b* \tmp* \the\sin*\a* \advance\ry*\tmp*
\raise\ry*\rlap{\kern\rx*\unhcopy\spl*}}

\def\calcnormal*#1{\b**=10000sp\a**\b**\k*\n*\advance\k*-\m*
\multiply\a**\k*\divide\a**\m*\a**=#1\a**\ifdim\a**<0pt\a**-\a**\fi
\ifdim\a**>\b**\d*=.96\a**\advance\d*.4\b**
\else\d*=.96\b**\advance\d*.4\a**\fi
\d*=.01\d*\r*\d*\divide\a**\r*\divide\b**\r*
\ifnum\k*<0\a**-\a**\fi\d*=#1\d*\ifdim\d*<0pt\b**-\b**\fi
\k*\a**\a**=\the\k*\dd*\k*\b**\b**=\the\k*\dd*}

\def\wavearcto#1(#2,#3)[#4]{\rlap{\toks0={#2}\toks1={#3}\relax
\calcnum*#1(#2,#3)\c*=\the\N*5sp\a*=.4\c*\a*=#4\a*\ifdim\a*<0pt\a*-\a*\fi
\advance\c*\a*\N*\c*\divide\N*20\multiply\N*2\advance\N*-1\multiply\N*4\relax
\d**=#1\Lengthunit\dd*=.012\d**
\divide\dd*\*ths \multiply\dd*\magnitude
\ifdim\d**<0pt\d**-\d**\fi\L*=.25\d**
\divide\d**\N*\divide\dd*\N*\L*=#4\L*\divide\L*\*ths
\m*\N*\divide\m*2\dm*=\the\m*0sp\l*\dm*
\sm*\n*\*one\loop\calcnormal*{#4}\calcparab*
\*shl*{1}\advance\n*\*one\calcparab*
\*shl*{1.3}\advance\n*\*one\calcparab*
\*shl*{1}\advance\n*2\dd*-\dd*\ifnum\n*<\N*\repeat\n*\N*\shl**{0pt}}}

\def\triangarcto#1(#2,#3)[#4]{\rlap{\toks0={#2}\toks1={#3}\relax
\calcnum*#1(#2,#3)\c*=\the\N*5sp\a*=.4\c*\a*=#4\a*\ifdim\a*<0pt\a*-\a*\fi
\advance\c*\a*\N*\c*\divide\N*20\multiply\N*2\advance\N*-1\multiply\N*2\relax
\d**=#1\Lengthunit\dd*=.012\d**
\divide\dd*\*ths \multiply\dd*\magnitude
\ifdim\d**<0pt\d**-\d**\fi\L*=.25\d**
\divide\d**\N*\divide\dd*\N*\L*=#4\L*\divide\L*\*ths
\m*\N*\divide\m*2\dm*=\the\m*0sp\l*\dm*
\sm*\n*\*one\loop\calcnormal*{#4}\calcparab*
\*shl*{1}\advance\n*2\dd*-\dd*\ifnum\n*<\N*\repeat\n*\N*\shl**{0pt}}}

\def\hr*#1{\L*=\xscale\Lengthunit\ifnum
\angle**=0\clap{\vrule width#1\L* height.1pt}\else
\L*=#1\L*\L*=.5\L*\rmov*(-\L*,0pt){\sm*}\rmov*(\L*,0pt){\sl*}\fi}

\def\shade#1[#2]{\rlap{\Lengthunit=#1\Lengthunit
\special{em:linewidth .001pt}\relax
\mov(0,#2.05){\hr*{.994}}\mov(0,#2.1){\hr*{.980}}\relax
\mov(0,#2.15){\hr*{.953}}\mov(0,#2.2){\hr*{.916}}\relax
\mov(0,#2.25){\hr*{.867}}\mov(0,#2.3){\hr*{.798}}\relax
\mov(0,#2.35){\hr*{.715}}\mov(0,#2.4){\hr*{.603}}\relax
\mov(0,#2.45){\hr*{.435}}\special{em:linewidth \the\linwid*}}}

\def\dshade#1[#2]{\rlap{\special{em:linewidth .001pt}\relax
\Lengthunit=#1\Lengthunit\if#2-\def\t*{+}\else\def\t*{-}\fi
\mov(0,\t*.025){\relax
\mov(0,#2.05){\hr*{.995}}\mov(0,#2.1){\hr*{.988}}\relax
\mov(0,#2.15){\hr*{.969}}\mov(0,#2.2){\hr*{.937}}\relax
\mov(0,#2.25){\hr*{.893}}\mov(0,#2.3){\hr*{.836}}\relax
\mov(0,#2.35){\hr*{.760}}\mov(0,#2.4){\hr*{.662}}\relax
\mov(0,#2.45){\hr*{.531}}\mov(0,#2.5){\hr*{.320}}\relax
\special{em:linewidth \the\linwid*}}}}

\def\vdot{\rlap{\kern-1.9pt\lower1.8pt\hbox{$\scriptstyle\bullet$}}}
\def\vtimes{\rlap{\kern-3pt\lower1.8pt\hbox{$\scriptstyle\times$}}}
\def\vDot{\rlap{\kern-2.3pt\lower2.7pt\hbox{$\bullet$}}}
\def\vTimes{\rlap{\kern-3.6pt\lower2.4pt\hbox{$\times$}}}

\def\arc(#1)[#2,#3]{{\k*=#2\l*=#3\m*=\l*
\advance\m*-6\ifnum\k*>\l*\relax\else
{\rotate(#2)\mov(#1,0){\sm*}}\loop
\ifnum\k*<\m*\advance\k*5{\rotate(\k*)\mov(#1,0){\sl*}}\repeat
{\rotate(#3)\mov(#1,0){\sl*}}\fi}}

\def\dasharc(#1)[#2,#3]{{\k**=#2\n*=#3\advance\n*-1\advance\n*-\k**
\L*=1000sp\L*#1\L* \multiply\L*\n* \multiply\L*\Nhalfperiods
\divide\L*57\N*\L* \divide\N*2000\ifnum\N*=0\N*1\fi
\r*\n*  \divide\r*\N* \ifnum\r*<2\r*2\fi
\m**\r* \divide\m**2 \l**\r* \advance\l**-\m** \N*\n* \divide\N*\r*
\k**\r* \multiply\k**\N* \dn*\n* \advance\dn*-\k** \divide\dn*2\advance\dn*\*one
\r*\l** \divide\r*2\advance\dn*\r* \advance\N*-2\k**#2\relax
\ifnum\l**<6{\rotate(#2)\mov(#1,0){\sm*}}\advance\k**\dn*
{\rotate(\k**)\mov(#1,0){\sl*}}\advance\k**\m**
{\rotate(\k**)\mov(#1,0){\sm*}}\loop
\advance\k**\l**{\rotate(\k**)\mov(#1,0){\sl*}}\advance\k**\m**
{\rotate(\k**)\mov(#1,0){\sm*}}\advance\N*-1\ifnum\N*>0\repeat
{\rotate(#3)\mov(#1,0){\sl*}}\else\advance\k**\dn*
\arc(#1)[#2,\k**]\loop\advance\k**\m** \r*\k**
\advance\k**\l** {\arc(#1)[\r*,\k**]}\relax
\advance\N*-1\ifnum\N*>0\repeat
\advance\k**\m**\arc(#1)[\k**,#3]\fi}}

\def\triangarc#1(#2)[#3,#4]{{\k**=#3\n*=#4\advance\n*-\k**
\L*=1000sp\L*#2\L* \multiply\L*\n* \multiply\L*\Nhalfperiods
\divide\L*57\N*\L* \divide\N*1000\ifnum\N*=0\N*1\fi
\d**=#2\Lengthunit \d*\d** \divide\d*57\multiply\d*\n*
\r*\n*  \divide\r*\N* \ifnum\r*<2\r*2\fi
\m**\r* \divide\m**2 \l**\r* \advance\l**-\m** \N*\n* \divide\N*\r*
\dt*\d* \divide\dt*\N* \dt*.5\dt* \dt*#1\dt*
\divide\dt*1000\multiply\dt*\magnitude
\k**\r* \multiply\k**\N* \dn*\n* \advance\dn*-\k** \divide\dn*2\relax
\r*\l** \divide\r*2\advance\dn*\r* \advance\N*-1\k**#3\relax
{\rotate(#3)\mov(#2,0){\sm*}}\advance\k**\dn*
{\rotate(\k**)\mov(#2,0){\sl*}}\advance\k**-\m**\advance\l**\m**\loop\dt*-\dt*
\d*\d** \advance\d*\dt*
\advance\k**\l**{\rotate(\k**)\rmov*(\d*,0pt){\sl*}}%
\advance\N*-1\ifnum\N*>0\repeat\advance\k**\m**
{\rotate(\k**)\mov(#2,0){\sl*}}{\rotate(#4)\mov(#2,0){\sl*}}}}

\def\wavearc#1(#2)[#3,#4]{{\k**=#3\n*=#4\advance\n*-\k**
\L*=4000sp\L*#2\L* \multiply\L*\n* \multiply\L*\Nhalfperiods
\divide\L*57\N*\L* \divide\N*1000\ifnum\N*=0\N*1\fi
\d**=#2\Lengthunit \d*\d** \divide\d*57\multiply\d*\n*
\r*\n*  \divide\r*\N* \ifnum\r*=0\r*1\fi
\m**\r* \divide\m**2 \l**\r* \advance\l**-\m** \N*\n* \divide\N*\r*
\dt*\d* \divide\dt*\N* \dt*.7\dt* \dt*#1\dt*
\divide\dt*1000\multiply\dt*\magnitude
\k**\r* \multiply\k**\N* \dn*\n* \advance\dn*-\k** \divide\dn*2\relax
\divide\N*4\advance\N*-1\k**#3\relax
{\rotate(#3)\mov(#2,0){\sm*}}\advance\k**\dn*
{\rotate(\k**)\mov(#2,0){\sl*}}\advance\k**-\m**\advance\l**\m**\loop\dt*-\dt*
\d*\d** \advance\d*\dt* \dd*\d** \advance\dd*1.3\dt*
\advance\k**\r*{\rotate(\k**)\rmov*(\d*,0pt){\sl*}}\relax
\advance\k**\r*{\rotate(\k**)\rmov*(\dd*,0pt){\sl*}}\relax
\advance\k**\r*{\rotate(\k**)\rmov*(\d*,0pt){\sl*}}\relax
\advance\k**\r*
\advance\N*-1\ifnum\N*>0\repeat\advance\k**\m**
{\rotate(\k**)\mov(#2,0){\sl*}}{\rotate(#4)\mov(#2,0){\sl*}}}}

\def\gmov*#1(#2,#3)#4{\rlap{\L*=#1\Lengthunit
\xL*=#2\L* \yL*=#3\L*
\rx* \gcos*\xL* \tmp* \gsin*\yL* \advance\rx*-\tmp*
\ry* \gcos*\yL* \tmp* \gsin*\xL* \advance\ry*\tmp*
\rx*=\xscale\rx* \ry*=\yscale\ry*
\xL* \the\cos*\rx* \tmp* \the\sin*\ry* \advance\xL*-\tmp*
\yL* \the\cos*\ry* \tmp* \the\sin*\rx* \advance\yL*\tmp*
\kern\xL*\raise\yL*\hbox{#4}}}

\def\rgmov*(#1,#2)#3{\rlap{\xL*#1\yL*#2\relax
\rx* \gcos*\xL* \tmp* \gsin*\yL* \advance\rx*-\tmp*
\ry* \gcos*\yL* \tmp* \gsin*\xL* \advance\ry*\tmp*
\rx*=\xscale\rx* \ry*=\yscale\ry*
\xL* \the\cos*\rx* \tmp* \the\sin*\ry* \advance\xL*-\tmp*
\yL* \the\cos*\ry* \tmp* \the\sin*\rx* \advance\yL*\tmp*
\kern\xL*\raise\yL*\hbox{#3}}}

\def\Earc(#1)[#2,#3][#4,#5]{{\k*=#2\l*=#3\m*=\l*
\advance\m*-6\ifnum\k*>\l*\relax\else\def\xscale{#4}\def\yscale{#5}\relax
{\angle**0\rotate(#2)}\gmov*(#1,0){\sm*}\loop
\ifnum\k*<\m*\advance\k*5\relax
{\angle**0\rotate(\k*)}\gmov*(#1,0){\sl*}\repeat
{\angle**0\rotate(#3)}\gmov*(#1,0){\sl*}\relax
\def\xscale{1}\def\yscale{1}\fi}}

\def\dashEarc(#1)[#2,#3][#4,#5]{{\k**=#2\n*=#3\advance\n*-1\advance\n*-\k**
\L*=1000sp\L*#1\L* \multiply\L*\n* \multiply\L*\Nhalfperiods
\divide\L*57\N*\L* \divide\N*2000\ifnum\N*=0\N*1\fi
\r*\n*  \divide\r*\N* \ifnum\r*<2\r*2\fi
\m**\r* \divide\m**2 \l**\r* \advance\l**-\m** \N*\n* \divide\N*\r*
\k**\r*\multiply\k**\N* \dn*\n* \advance\dn*-\k** \divide\dn*2\advance\dn*\*one
\r*\l** \divide\r*2\advance\dn*\r* \advance\N*-2\k**#2\relax
\ifnum\l**<6\def\xscale{#4}\def\yscale{#5}\relax
{\angle**0\rotate(#2)}\gmov*(#1,0){\sm*}\advance\k**\dn*
{\angle**0\rotate(\k**)}\gmov*(#1,0){\sl*}\advance\k**\m**
{\angle**0\rotate(\k**)}\gmov*(#1,0){\sm*}\loop
\advance\k**\l**{\angle**0\rotate(\k**)}\gmov*(#1,0){\sl*}\advance\k**\m**
{\angle**0\rotate(\k**)}\gmov*(#1,0){\sm*}\advance\N*-1\ifnum\N*>0\repeat
{\angle**0\rotate(#3)}\gmov*(#1,0){\sl*}\def\xscale{1}\def\yscale{1}\else
\advance\k**\dn* \Earc(#1)[#2,\k**][#4,#5]\loop\advance\k**\m** \r*\k**
\advance\k**\l** {\Earc(#1)[\r*,\k**][#4,#5]}\relax
\advance\N*-1\ifnum\N*>0\repeat
\advance\k**\m**\Earc(#1)[\k**,#3][#4,#5]\fi}}

\def\triangEarc#1(#2)[#3,#4][#5,#6]{{\k**=#3\n*=#4\advance\n*-\k**
\L*=1000sp\L*#2\L* \multiply\L*\n* \multiply\L*\Nhalfperiods
\divide\L*57\N*\L* \divide\N*1000\ifnum\N*=0\N*1\fi
\d**=#2\Lengthunit \d*\d** \divide\d*57\multiply\d*\n*
\r*\n*  \divide\r*\N* \ifnum\r*<2\r*2\fi
\m**\r* \divide\m**2 \l**\r* \advance\l**-\m** \N*\n* \divide\N*\r*
\dt*\d* \divide\dt*\N* \dt*.5\dt* \dt*#1\dt*
\divide\dt*1000\multiply\dt*\magnitude
\k**\r* \multiply\k**\N* \dn*\n* \advance\dn*-\k** \divide\dn*2\relax
\r*\l** \divide\r*2\advance\dn*\r* \advance\N*-1\k**#3\relax
\def\xscale{#5}\def\yscale{#6}\relax
{\angle**0\rotate(#3)}\gmov*(#2,0){\sm*}\advance\k**\dn*
{\angle**0\rotate(\k**)}\gmov*(#2,0){\sl*}\advance\k**-\m**
\advance\l**\m**\loop\dt*-\dt* \d*\d** \advance\d*\dt*
\advance\k**\l**{\angle**0\rotate(\k**)}\rgmov*(\d*,0pt){\sl*}\relax
\advance\N*-1\ifnum\N*>0\repeat\advance\k**\m**
{\angle**0\rotate(\k**)}\gmov*(#2,0){\sl*}\relax
{\angle**0\rotate(#4)}\gmov*(#2,0){\sl*}\def\xscale{1}\def\yscale{1}}}

\def\waveEarc#1(#2)[#3,#4][#5,#6]{{\k**=#3\n*=#4\advance\n*-\k**
\L*=4000sp\L*#2\L* \multiply\L*\n* \multiply\L*\Nhalfperiods
\divide\L*57\N*\L* \divide\N*1000\ifnum\N*=0\N*1\fi
\d**=#2\Lengthunit \d*\d** \divide\d*57\multiply\d*\n*
\r*\n*  \divide\r*\N* \ifnum\r*=0\r*1\fi
\m**\r* \divide\m**2 \l**\r* \advance\l**-\m** \N*\n* \divide\N*\r*
\dt*\d* \divide\dt*\N* \dt*.7\dt* \dt*#1\dt*
\divide\dt*1000\multiply\dt*\magnitude
\k**\r* \multiply\k**\N* \dn*\n* \advance\dn*-\k** \divide\dn*2\relax
\divide\N*4\advance\N*-1\k**#3\def\xscale{#5}\def\yscale{#6}\relax
{\angle**0\rotate(#3)}\gmov*(#2,0){\sm*}\advance\k**\dn*
{\angle**0\rotate(\k**)}\gmov*(#2,0){\sl*}\advance\k**-\m**
\advance\l**\m**\loop\dt*-\dt*
\d*\d** \advance\d*\dt* \dd*\d** \advance\dd*1.3\dt*
\advance\k**\r*{\angle**0\rotate(\k**)}\rgmov*(\d*,0pt){\sl*}\relax
\advance\k**\r*{\angle**0\rotate(\k**)}\rgmov*(\dd*,0pt){\sl*}\relax
\advance\k**\r*{\angle**0\rotate(\k**)}\rgmov*(\d*,0pt){\sl*}\relax
\advance\k**\r*
\advance\N*-1\ifnum\N*>0\repeat\advance\k**\m**
{\angle**0\rotate(\k**)}\gmov*(#2,0){\sl*}\relax
{\angle**0\rotate(#4)}\gmov*(#2,0){\sl*}\def\xscale{1}\def\yscale{1}}}

\newcount\CatcodeOfAtSign
\CatcodeOfAtSign=\the\catcode`\@
\catcode`\@=11
\def\@arc#1[#2][#3]{\rlap{\Lengthunit=#1\Lengthunit
\sm*\l*arc(#2.1914,#3.0381)[#2][#3]\relax
\mov(#2.1914,#3.0381){\l*arc(#2.1622,#3.1084)[#2][#3]}\relax
\mov(#2.3536,#3.1465){\l*arc(#2.1084,#3.1622)[#2][#3]}\relax
\mov(#2.4619,#3.3086){\l*arc(#2.0381,#3.1914)[#2][#3]}}}

\def\dash@arc#1[#2][#3]{\rlap{\Lengthunit=#1\Lengthunit
\d*arc(#2.1914,#3.0381)[#2][#3]\relax
\mov(#2.1914,#3.0381){\d*arc(#2.1622,#3.1084)[#2][#3]}\relax
\mov(#2.3536,#3.1465){\d*arc(#2.1084,#3.1622)[#2][#3]}\relax
\mov(#2.4619,#3.3086){\d*arc(#2.0381,#3.1914)[#2][#3]}}}

\def\wave@arc#1[#2][#3]{\rlap{\Lengthunit=#1\Lengthunit
\w*lin(#2.1914,#3.0381)\relax
\mov(#2.1914,#3.0381){\w*lin(#2.1622,#3.1084)}\relax
\mov(#2.3536,#3.1465){\w*lin(#2.1084,#3.1622)}\relax
\mov(#2.4619,#3.3086){\w*lin(#2.0381,#3.1914)}}}

\def\bezier#1(#2,#3)(#4,#5)(#6,#7){\N*#1\l*\N* \advance\l*\*one
\d* #4\Lengthunit \advance\d* -#2\Lengthunit \multiply\d* \*two
\b* #6\Lengthunit \advance\b* -#2\Lengthunit
\advance\b*-\d* \divide\b*\N*
\d** #5\Lengthunit \advance\d** -#3\Lengthunit \multiply\d** \*two
\b** #7\Lengthunit \advance\b** -#3\Lengthunit
\advance\b** -\d** \divide\b**\N*
\mov(#2,#3){\sm*{\loop\ifnum\m*<\l*
\a*\m*\b* \advance\a*\d* \divide\a*\N* \multiply\a*\m*
\a**\m*\b** \advance\a**\d** \divide\a**\N* \multiply\a**\m*
\rmov*(\a*,\a**){\unhcopy\spl*}\advance\m*\*one\repeat}}}

\catcode`\*=12

\newcount\n@ast
\def\n@ast@#1{\n@ast0\relax\get@ast@#1\end}
\def\get@ast@#1{\ifx#1\end\let\next\relax\else
\ifx#1*\advance\n@ast1\fi\let\next\get@ast@\fi\next}

\newif\if@up \newif\if@dwn
\def\up@down@#1{\@upfalse\@dwnfalse
\if#1u\@uptrue\fi\if#1U\@uptrue\fi\if#1+\@uptrue\fi
\if#1d\@dwntrue\fi\if#1D\@dwntrue\fi\if#1-\@dwntrue\fi}

\def\halfcirc#1(#2)[#3]{{\Lengthunit=#2\Lengthunit\up@down@{#3}\relax
\if@up\mov(0,.5){\@arc[-][-]\@arc[+][-]}\fi
\if@dwn\mov(0,-.5){\@arc[-][+]\@arc[+][+]}\fi
\def\lft{\mov(0,.5){\@arc[-][-]}\mov(0,-.5){\@arc[-][+]}}\relax
\def\rght{\mov(0,.5){\@arc[+][-]}\mov(0,-.5){\@arc[+][+]}}\relax
\if#3l\lft\fi\if#3L\lft\fi\if#3r\rght\fi\if#3R\rght\fi
\n@ast@{#1}\relax
\ifnum\n@ast>0\if@up\shade[+]\fi\if@dwn\shade[-]\fi\fi
\ifnum\n@ast>1\if@up\dshade[+]\fi\if@dwn\dshade[-]\fi\fi}}

\def\halfdashcirc(#1)[#2]{{\Lengthunit=#1\Lengthunit\up@down@{#2}\relax
\if@up\mov(0,.5){\dash@arc[-][-]\dash@arc[+][-]}\fi
\if@dwn\mov(0,-.5){\dash@arc[-][+]\dash@arc[+][+]}\fi
\def\lft{\mov(0,.5){\dash@arc[-][-]}\mov(0,-.5){\dash@arc[-][+]}}\relax
\def\rght{\mov(0,.5){\dash@arc[+][-]}\mov(0,-.5){\dash@arc[+][+]}}\relax
\if#2l\lft\fi\if#2L\lft\fi\if#2r\rght\fi\if#2R\rght\fi}}

\def\halfwavecirc(#1)[#2]{{\Lengthunit=#1\Lengthunit\up@down@{#2}\relax
\if@up\mov(0,.5){\wave@arc[-][-]\wave@arc[+][-]}\fi
\if@dwn\mov(0,-.5){\wave@arc[-][+]\wave@arc[+][+]}\fi
\def\lft{\mov(0,.5){\wave@arc[-][-]}\mov(0,-.5){\wave@arc[-][+]}}\relax
\def\rght{\mov(0,.5){\wave@arc[+][-]}\mov(0,-.5){\wave@arc[+][+]}}\relax
\if#2l\lft\fi\if#2L\lft\fi\if#2r\rght\fi\if#2R\rght\fi}}

\catcode`\*=11

\def\Circle#1(#2){\halfcirc#1(#2)[u]\halfcirc#1(#2)[d]\n@ast@{#1}\relax
\ifnum\n@ast>0\L*=\xscale\Lengthunit
\ifnum\angle**=0\clap{\vrule width#2\L* height.1pt}\else
\L*=#2\L*\L*=.5\L*\special{em:linewidth .001pt}\relax
\rmov*(-\L*,0pt){\sm*}\rmov*(\L*,0pt){\sl*}\relax
\special{em:linewidth \the\linwid*}\fi\fi}

\catcode`\*=12

\def\wavecirc(#1){\halfwavecirc(#1)[u]\halfwavecirc(#1)[d]}

\def\dashcirc(#1){\halfdashcirc(#1)[u]\halfdashcirc(#1)[d]}

\def\xscale{1}
\def\yscale{1}

\def\Ellipse#1(#2)[#3,#4]{\def\xscale{#3}\def\yscale{#4}\relax
\Circle#1(#2)\def\xscale{1}\def\yscale{1}}

\def\dashEllipse(#1)[#2,#3]{\def\xscale{#2}\def\yscale{#3}\relax
\dashcirc(#1)\def\xscale{1}\def\yscale{1}}

\def\waveEllipse(#1)[#2,#3]{\def\xscale{#2}\def\yscale{#3}\relax
\wavecirc(#1)\def\xscale{1}\def\yscale{1}}

\def\halfEllipse#1(#2)[#3][#4,#5]{\def\xscale{#4}\def\yscale{#5}\relax
\halfcirc#1(#2)[#3]\def\xscale{1}\def\yscale{1}}

\def\halfdashEllipse(#1)[#2][#3,#4]{\def\xscale{#3}\def\yscale{#4}\relax
\halfdashcirc(#1)[#2]\def\xscale{1}\def\yscale{1}}

\def\halfwaveEllipse(#1)[#2][#3,#4]{\def\xscale{#3}\def\yscale{#4}\relax
\halfwavecirc(#1)[#2]\def\xscale{1}\def\yscale{1}}

\catcode`\@=\the\CatcodeOfAtSign

\section{Introduction}

The investigation of the energy spectra of hydrogenic atoms is of great
importance for high accuracy verification of the Standard Model and
derivation more correct values of fundamental physical constants (the fine
structure constant, the masses of the muon and electron, the proton
charge radius etc.) \cite{EGS,Sokolov,MT}. In the last few years
significant interest in this research area is connected with the atom
of muonic hydrogen \cite{K1,K2,B1,B2}. It has motivated by the activity
of the experimental study of the $(2P-2S)$ Lamb shift and hyperfine structure
in muonic hydrogen. The measurement of the $(2P-2S)$ Lamb shift with the
accuracy 30 $ppm$ in the $(\mu p)$ would allow us to obtain the value
of the proton charge radius with a precision $10^{-3}$ which is an order
of the magnitude better in the comparison with different methods including
the elastic electron-proton scattering and the value of the $(2P-2S)$
Lamb shift in the atom of electronic hydrogen. The measurement of the ground
state hyperfine splitting in muonic hydrogen with a similar accuracy
would allow to determine the value of the Zemach radius \cite{Zemach} with
a precision $10^{-3}$ \cite{FM2004,apm} which can be considered as a new
fundamental parameter regarding to the hydrogen atom.
Then it can be used in the calculation of new theoretical
quantity of the hyperfine splitting in electronic hydrogen and the restriction
on the value of the proton polarizability contribution \cite{DS,BE,CFM,Pineda}.

There exist another experimental task of the investigation of large fine
structure interval $(1S-2S)$ in muonic hydrogen and the muonic hydrogen - muonic
deuterium isotope shift for this splitting \cite{Breit,theory}. It would
allow to obtain new data about the proton and deuteron charge radii.
It should be pointed out that both marked intervals are among the most
precise measured quantities for the atom of electronic hydrogen.
Experimental accuracy for the measurement of the hydrogen-deuterium isotope
shift for the transition $(1S-2S)$ increased by three orders of the magnitude
during last ten years. At present time it is equal \cite{HUG}
\begin{equation}
\Delta \nu_{IS}=[E(2S)-E(1S)]_D-[E(2S)-E(1S)]_H=670~994~334.64(15)~kHz.
\end{equation}
The interval $(1S-2S)$ for the hydrogen atom was measured with extremely
high accuracy of order of $10^{-2}$ kHz \cite{Udem}:
\begin{equation}
\Delta\nu_{1S-2S}(H)=2~466~061~413~187~103(46)~Hz,~~~\delta=1.8\cdot 10^{-14}.
\end{equation}
Experimental investigations of the energy intervals (1) and (2) in muonic
hydrogen are on the stage of preliminary preparation.

Theoretical studies of various contributions to the energy levels of
muonic atoms were made long ago \cite{MS,BR,KP} (see also the Refs. in the
review article \cite{EGS}). In the last years the calculation of different
order corrections in the energy spectrum of muonic hydrogen was connected
for the most part with the $(2P-2S)$ Lamb shift and hyperfine structure of
$S$-states \cite{FM3,FM,CFM,KP1999,VP2004}. In these papers the two-particle
interaction operator was constructed which gave the contributions of orders
$\alpha^5$ and $\alpha^6$ to the interval $(2P-2S)$ and hyperfine splitting
of the $1S$ and $2S$ states. At present the necessity exists for theoretical
investigation of the corrections of orders $\alpha^3$, $\alpha^4$ and
$\alpha^5$ to the $1S$ and $2S$ Lamb shift in muonic hydrogen and muonic
deuterium, the isotope shift $(\mu p)$ - $(\mu d)$ for the transition
$(1S-2S)$ and the fine structure interval $[E(1S)-8E(2S)]$ which are yet
unknown. Such calculations can initiate the experimental investigations
of the fine structure intervals (1) and (2) in muonic hydrogen in order
to obtain more exact values of such fundamental physical constants as the
proton and deuteron charge radii and the muon mass.

In this paper we have derived numerical results for the contributions of orders
$\alpha^3$, $\alpha^4$ and $\alpha^5$ to the splitting $(1S-2S)$
and the isotope shift $(\mu p)$ - $(\mu d)$ for this splitting.
In doing so, we obtained numerical values for some contributions on the basis
of known analytical expressions. The most part of the corrections for the energy
levels $1S$ and $2S$ and the isotope shift in muonic hydrogen is determined
originally in the integral form which then is used for numerical estimations.
The dependence on principal quantum number $n$ for the most part of the
contributions is not trivial that is not reduces to the factor $1/n^3$.
The reason has to do with characteristic value of the particle momenta and
will be discussed below. The aim of this work consists in the calculation
of quantum electrodynamic corrections in the muonic hydrogen - muonic
deuterium isotope shift for the transition $(1S-2S)$ and the fine structure
interval $[E(1S)-8E(2S)]$ both for $(\mu p)$ and $(\mu d)$ with the accuracy
$10^{-9}$ which can be considered as a proper guide for the future experiments
and the retrieval of the information about the proton and deuteron charge radii and
the muon mass.

Fine structure of the energy levels of hydrogenic atoms has been studied
for a long time on the basis of different approaches \cite{ES,ES1,theory,EGS}.
For the case of $S$-states of hydrogen-like atoms consisting of the particles
with masses $m_1$ and $m_2$ it can be written with the accuracy $O((Z\alpha)^4)$
as follows:
\begin{equation}
E_n=m_1+m_2-\frac{\mu^2(Z\alpha)^2}{2n^2}-\frac{\mu(Z\alpha)^4}{2n^3}\left[
1-\frac{3}{4n}+\frac{\mu^2}{4m_1m_2n}\right]=
\end{equation}
\begin{displaymath}
=\Biggl\{{{\mu p(1S):~1~043~927~826~470.3586~meV;~\mu p(2S):~1~043~929~722~866.0601~meV}\atop
{\mu d(1S):~1~981~268~455~762.7537~meV;~\mu d(2S):~1~981~270~453~188.8081~meV}}.
\end{displaymath}
The error of relative order in Eq.(3) which is determined by the uncertainties
of the fine structure constant $\alpha$ and the particle masses reaches
the value $10^{-7}$. Nevertheless, we represented the numerical values in Eq.(3)
with the precision 0.0001 meV because it is important when we consider various
fine structure intervals in the energy spectrum. The following values of
fundamental physical constants are used: $\alpha^{-1}=137.03599976(50)$,
$m_\mu$ = 0.105658357 (5) GeV, $m_p$ = 0.938271998 (38) GeV,
$m_d$ = 1.875612762 (75) GeV \cite{MT}. The contribution of Eq.(3) to the
$(\mu p)$ - $(\mu d)$ isotope shift for the splitting $(1S-2S)$ is the crucial
(see the Table I). In addition, there exist a number of important effects
both electromagnetic and strong interactions which must be taken into account
for the determination of the isotope shift value with the accuracy up to terms
of order $\alpha^5$.

\section{Effects of one-loop and two-loop vacuum polarization in
one-photon interaction}

Our calculations of various energy levels of hydrogen-like atoms are
performed within the framework of the quasipotential approach in which
a bound state of two particles is described by the Shroedinger-type
equation \cite{MF85,FM1}:
\begin{eqnarray}
\left[G^f\right]^{-1}\psi_M\equiv\left(\frac{b^2}{2\mu_R}-\frac{{\bf p}^2}{2\mu_R}\right)
\psi_M({\bf p})=
\int\frac{d{\bf q}}{(2\pi)^3}V({\bf p},{\bf q},M)\psi_M({\bf q}),
\end{eqnarray}
where
\begin{displaymath}
b^2=E_1^2-m_1^2=E_2^2-m_2^2,
\end{displaymath}
$\mu_R=E_1E_2/M$ is the relativistic reduced mass, $M=E_1+E_2$ is the mass
of the bound state. The quasipotential in Eq.(4) is constructed in quantum
electrodynamics by the perturbation theory with the use of the two-particle
scattering amplitude $T$ projected onto the positive frequency states outside
the mass surface at zero relative energies of the particles:
\begin{equation}
V=V^{(1)}+V^{(2)}+V^{(3)}+...,~~~~~T=T^{(1)}+T^{(2)}+T^{(3)}+...,
\end{equation}
\begin{equation}
V^{(1)}=T^{(1)},~V^{(2)}=T^{(2)}-T^{(1)}\times G^f\times T^{(1)}, ...~~.
\end{equation}

\begin{figure}[t!]
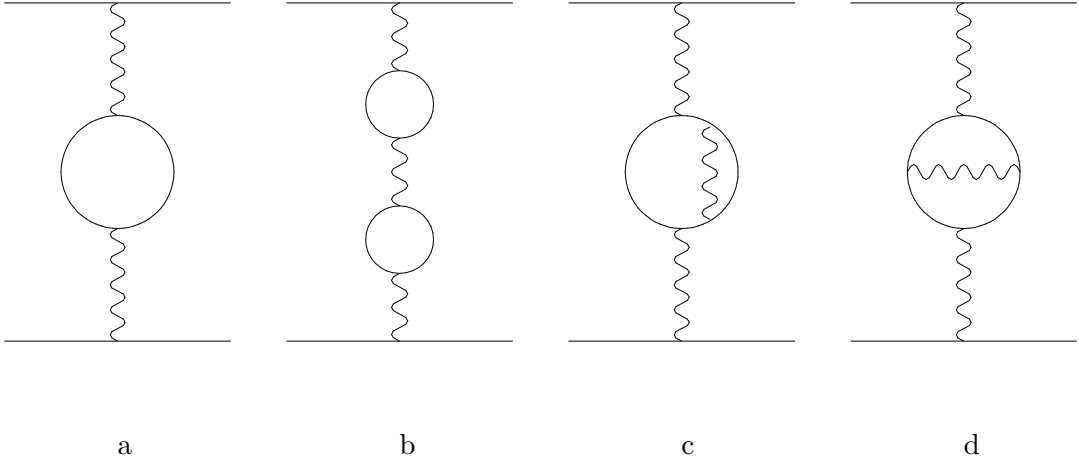

\magnitude=2000
\GRAPH(hsize=15){
\mov(0,0){\lin(2,0)}%
\mov(0,3){\lin(2,0)}%
\mov(2.5,0){\lin(2,0)}%
\mov(2.5,3){\lin(2,0)}%
\mov(5,0){\lin(2,0)}%
\mov(5,3){\lin(2,0)}%
\mov(7.5,0){\lin(2,0)}%
\mov(7.5,3){\lin(2,0)}%
\mov(1,1.5){\Circle(1.)}%
\mov(1,0){\wavelin(0,1)}%
\mov(1,2){\wavelin(0,1)}%
\mov(3.5,0){\wavelin(0,0.6)}%
\mov(3.5,1.2){\wavelin(0,0.6)}%
\mov(3.5,2.4){\wavelin(0,0.6)}%
\mov(3.5,0.9){\Circle(0.6)}%
\mov(3.5,2.1){\Circle(0.6)}%
\mov(6,0){\wavelin(0,1)}%
\mov(6,2){\wavelin(0,1)}%
\mov(6,1.5){\Circle(1.)}%
\mov(6.25,1.08){\wavelin(0,0.82)}%
\mov(8.5,0){\wavelin(0,1.)}%
\mov(8.5,2.){\wavelin(0,1)}%
\mov(8.,1.5){\wavelin(1,0)}%
\mov(8.5,1.5){\Circle(1.0)}%
\mov(1.,-1.){a}%
\mov(3.5,-1.){b}%
\mov(6.,-1.){c}%
\mov(8.5,-1.){d}%
}
\caption{Effects of one-loop and two-loop vacuum polarization
in one-photon interaction.}
\end{figure}

The initial approximation of the quasipotential $V({\bf p},{\bf q},M)$
for a bound system was selected in the form of the usual Coulomb potential.
The increase in the lepton mass in muonic hydrogen compared with its
electronic counterpart decreases the radius of the Bohr orbit in the $(\mu p)$.
As a result, the Compton wave length of the electron and the radius of the
Bohr orbit become commensurable:
\begin{equation}
\frac{\hbar^2}{\mu e^2}:\frac{\hbar}{m_ec}=0.737384,
\end{equation}
where $m_e$ is the mass of the electron, $\mu$ is the reduced mass in the
$(\mu p)$ atom. This substantially enhances the role played by vacuum
polarization effects in the energy spectrum of muonic hydrogen \cite{t4}.
Corrections of the one-loop and two-loop vacuum polarization to the
quasipotential of one-photon interaction are shown in Fig.1.

To determine the contribution of diagram (a) in Fig.1 (electronic vacuum
polarization) to the particle interaction operator we must perform the
following substitution in the photon propagator \cite{t4}:
\begin{equation}
\frac{1}{k^2}\to \frac{\alpha}{3\pi}\int_1^\infty ds
\frac{\sqrt{s^2-1}(2s^2+1)}{s^4(k^2+4m_e^2s^2)}.
\end{equation}
If
\begin{displaymath}
(-k^2)={\bf k}^2\sim\mu_e^2(Z\alpha)^2\sim m_e^2(Z\alpha)^2
\end{displaymath}
(electronic hydrogen, $\mu_e$ is the reduced mass of two particles in
the hydrogen atom), then, ignoring the first term in the denominator
in the right-hand side of Eq.(8) we obtain:
\begin{displaymath}
-\alpha/15\pi m_e^2.
\end{displaymath}
However, if
\begin{displaymath}
{\bf k}^2\sim \mu^2(Z\alpha)^2\sim m_1^2(Z\alpha)^2
\end{displaymath}
(muonic hydrogen, $m_1$ is the mass of the muon), then $\mu \alpha$ and $m_e$
are values of one order and we cannot expand the denominator in Eq.(8) in
$\alpha$. With muonic hydrogen, we must construct the particle interaction
operator in the one-photon approximation using exact expression (8). Further,
we take into account that the electron vacuum polarization gives the
contributions of orders $\alpha^3$, $\alpha^4$ and $\alpha^5$ to the energy
spectrum of $S$-states in the $(\mu p)$ atom.

The modification of the Coulomb potential
\begin{displaymath}
V^C({\bf k})=-Ze^2/{\bf k}^2
\end{displaymath}
caused by the vacuum polarization
(VP) is determined taking into account (8) by the following expression
in the momentum representation \cite{t4}:
\begin{equation}
V^C_{VP}({\bf k})=-4\pi Z\alpha\frac{\alpha}{\pi}\int_1^\infty\frac{\sqrt
{\xi^2-1}}{3\xi^4}\frac{(2\xi^2+1)}{{\bf k}^2+4m_e^2\xi^2}d\xi
\end{equation}
The Fourier transform of Eq.(9) gives the corresponding operator in the
coordinate representation:
\begin{equation}
V^C_{VP}(r)=\frac{\alpha}{3\pi}\int_1^\infty d\xi\frac{\sqrt{\xi^2-1}(
2\xi^2+1)}{\xi^4}\left(-\frac{Z\alpha}{r}e^{-2m_e\xi r}\right).
\end{equation}
The potential (10) allows us to obtain the correction of the electron
vacuum polarization of order $\alpha^3$ to the energy levels of $1S$ and
$2S$-states in muonic hydrogen. Accounting exact expressions of the wave
functions for $1S$ and $2S$-states
\begin{equation}
\psi_{100}(r)=\frac{W^{3/2}}{\sqrt{\pi}}e^{-Wr},~~
\psi_{200}(r)=\frac{W^{3/2}}{2\sqrt{2\pi}}e^{-Wr/2}\left
(1-\frac{Wr}{2}\right),~~~W=\mu Z\alpha,
\end{equation}
we represent this correction in the form:
\begin{equation}
\Delta E_{1\gamma,VP}(1S)=-\frac{\mu(Z\alpha)^2\alpha}{3\pi}\int_1^\infty
\rho(\xi)d\xi\frac{1}{p^2_1(\xi)},~p_1(\xi)=1+\frac{m_e\xi}{W},
\rho(\xi)=\frac{\sqrt{\xi^2-1}(2\xi^2+1)}{\xi^4},
\end{equation}
\begin{equation}
\Delta E_{1\gamma,VP}(2S)=-\frac{\mu(Z\alpha)^2\alpha}{6\pi}\int_1^\infty
\rho(\xi)d\xi\left(\frac{1}{p^2_2(\xi)}-\frac{2}{p_2^3(\xi)}+\frac{3}
{2p_2^4(\xi)}\right),~p_2(\xi)=1+\frac{2m_e\xi}{W}.
\end{equation}
The electron vacuum polarization effects are very sensitive to the bound
state structure because the characteristic momentum of the particles in
muonic hydrogen atom is $(\mu Z\alpha)$. So, the contribution of the
vacuum polarization amplitudes in one-photon interaction cannot be expressed
by a simple factor $|\psi^C(0)|^2\sim 1/n^3$ but it is dependent on principal
quantum number $n$ in more complicated form which is represented by
expressions (12) and (13). Numerical values of the electron vacuum polarization
contribution to the $1S$ and $2S$ Lamb shift in muonic hydrogen and muonic
deuterium differ only due to reduced mass of two particles:
\begin{equation}
\Delta E_{1\gamma,VP}=
\Biggl\{{{\mu p(1S):~-1898.8379~meV;~\mu p(2S):~-219.5849~meV}\atop
{\mu d(1S):~-2129.2820~meV;~\mu d(2S):~-245.3205~meV}}
\end{equation}
The contribution of the muon vacuum polarization (MVP) can be found also
with the use of Eq.(9) in which the replacement $m_e\to m_1$ must be carried
out. This correction of order $\alpha^5$ to the energy spectrum is included
in Table I together with the muon self energy correction (MSE).

\begin{figure}[t!]
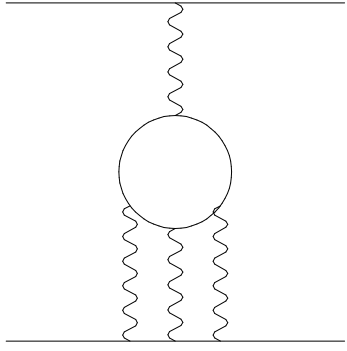

\magnitude=2000
\GRAPH(hsize=15){
\mov(4,0){\lin(3,0)}%
\mov(4,3){\lin(3,0)}%
\mov(5.5,1.5){\Circle(1.)}%
\mov(5.5,0.){\wavelin(0,1.)}%
\mov(5.9,0){\wavelin(0,1.2)}%
\mov(5.1,0){\wavelin(0,1.2)}%
\mov(5.5,3.){\wavelin(0,-1.)}%
}
\caption{The Wichmann-Kroll correction. The wave line denotes the Coulomb
photon.}
\end{figure}

The diagram in Fig.2 has the form of one-loop vacuum polarization. It
determines the correction to the energy spectrum of order $\alpha^5$ known
as the Wichmann-Kroll correction \cite{WK,MPS}. The corresponding potential
was represented in the integral form:
\begin{equation}
\Delta V^{WK}(r)=\frac{\alpha(Z\alpha)^3}{\pi r}\int_0^\infty \frac{d\zeta}
{\zeta^4} e^{-2m_e\zeta r}\left[-\frac{\pi^2}{12}\sqrt{\zeta^2-1}\theta(\zeta-1)
+\int_0^\zeta dx\sqrt{\zeta^2-x^2} f^{WK}(x)\right].
\end{equation}
Exact expression for the spectral function $f^{WK}$ is written in Ref.\cite{EGS,WK,MPS}.
Averaging the potential (15) over the Coulomb wave functions (11) we
expressed the corrections to the $S$-states as follows:
\begin{equation}
\Delta E^{WK}(1S)=\frac{\mu\alpha(Z\alpha)^4}{\pi}
\int_0^\infty \frac{d\zeta}{\zeta^4p_1^2(\zeta)} \left[-\frac{\pi^2}{12}
\sqrt{\zeta^2-1}\theta(\zeta-1)+\int_0^\zeta dx\sqrt{\zeta^2-x^2} f^{WK}(x)
\right],
\end{equation}
\begin{equation}
\Delta E^{WK}(2S)=\frac{\mu\alpha(Z\alpha)^4}{2\pi}
\int_0^\infty \frac{d\zeta}{\zeta^4}\left(\frac{1}{p_2^2(\zeta)}-\frac{2}
{p_2^3(\zeta)}+\frac{3}{2p_2^4(\zeta)}\right)\times
\end{equation}
\begin{displaymath}
\times\left[-\frac{\pi^2}{12}\sqrt{\zeta^2-1}\theta(\zeta-1)+\int_0^\zeta dx\sqrt{\zeta^2-x^2} f^{WK}(x)
\right].
\end{displaymath}
Numerical values of Eqs. (16) and (17) are included in Table I.

Let us consider the modification of the Coulomb potential by virtue of
two-loop vacuum polarization (see Fig.1(b,c,d)). The contribution of the
diagram (b) in Fig.1 containing two sequential electronic loops can be
obtained by applying the replacement (8) two times in the photon propagator.
In the coordinate representation the required interaction operator of the
particles has the form:
\begin{equation}
V_{1\gamma,~VP-VP}^C(r)=\frac{\alpha^2}{9\pi^2}
\int_1^\infty\rho(\xi)d\xi\int_1^\infty\rho(\eta)d\eta\left(-\frac{Z\alpha}{r}
\right)\frac{1}{(\xi^2-\eta^2)}\left(\xi^2e^{-2m_e\xi r}-\eta^2e^{-2m_e\eta r}
\right).
\end{equation}
It gives the following results for the energy spectrum:
\begin{equation}
\Delta E_{1\gamma,~VP-VP}(1S)=-\frac{\mu\alpha^2(Z\alpha)^2}{9\pi^2}
\int_1^\infty\rho(\xi)d\xi\int_1^\infty\rho(\eta)d\eta\frac{1}{(\xi^2-\eta^2)}
\left(\frac{\xi^2}{p_1^2(\xi)}-\frac{\eta^2}{p_1^2(\eta)}\right)=
\end{equation}
\begin{displaymath}
=\Biggl\{{{\mu p:~-1.8816~meV}\atop
{\mu d:~-2.1871~meV}},
\end{displaymath}
\begin{equation}
\Delta E_{1\gamma,~VP-VP}(2S)=-\frac{\mu\alpha^2(Z\alpha)^2}{18\pi^2}
\int_1^\infty\rho(\xi)d\xi\int_1^\infty\rho(\eta)d\eta\frac{1}
{(\xi^2-\eta^2)}\times
\end{equation}
\begin{displaymath}
\times\left[\xi^2\left(\frac{1}{p_2^2(\xi)}-\frac{2}{p_2^3(\xi)}+\frac{3}{2p_2^4(\xi)}
\right)-\eta^2\left(\frac{1}{p_2^2(\eta)}-\frac{2}{p_2^3(\eta)}+
\frac{3}{2p_2^4(\eta)}\right)\right]=\Biggl\{{{\mu p:~-0.2426~meV}\atop
{\mu d:~-0.2811~meV}}.
\end{displaymath}

The contributions of the diagrams (b,c) in Fig.1 which are determined by
the second order polarization operator can be calculated after the following
replacement in the photon propagator \cite{EGS1}:
\begin{equation}
\frac{1}{k^2}\to \left(\frac{\alpha}{\pi}\right)^2\int_0^1\frac{f(v)}
{4m_e^2+k^2(1-v^2)}dv=\left(\frac{\alpha}{\pi}\right)^2\frac{2}{3}\int_0^1 dv
\frac{v}{4m_e^2+k^2(1-v^2)}\times
\end{equation}
\begin{displaymath}
\times\Biggl\{(3-v^2)(1+v^2)\left[Li_2\left(-\frac{1-v}{1+v}\right)+2Li_2
\left(\frac{1-v}{1+v}\right)+\frac{3}{2}\ln\frac{1+v}{1-v}\ln\frac{1+v}{2}-
\ln\frac{1+v}{1-v}\ln v\right]+
\end{displaymath}
\begin{displaymath}
\left[\frac{11}{16}(3-v^2)(1+v^2)+\frac{v^4}{4}\right]\ln\frac{1+v}{1-v}+
\left[\frac{3}{2}v(3-v^2)\ln\frac{1-v^2}{4}-2v(3-v^2)\ln v\right]+
\frac{3}{8}v(5-3v^2)\Biggr\}.
\end{displaymath}
The contribution value can then conveniently be calculated in the coordinate
representation with the reduction of the interparticle interaction potential
to the form:
\begin{equation}
\Delta V_{1\gamma,~ 2-loop~VP}^C=-\frac{2}{3}\frac{Z\alpha}{r}
\left(\frac{\alpha}{\pi}\right)^2\int_0^1\frac{f(v)dv}{(1-v^2)}
e^{-\frac{2m_er}{\sqrt{1-v^2}}}.
\end{equation}
The operator (22) gives the following corrections to the Lamb shift of
$S$-levels in muonic hydrogen and deuterium:
\begin{equation}
\Delta E_{1\gamma,~2-loop~VP}(1S)=-\frac{2}{3\pi^2}\mu\alpha^2(Z\alpha)^2
\frac{W^2}{m_e^2}\int_0^1\frac{f(v)dv}{\left(1+\frac{W\sqrt{1-v^2}}{m_e}
\right)^2}=
\Biggl\{{{\mu p:~-12.6144~meV}\atop{\mu d:~-14.0141~meV}}
\end{equation}
\begin{equation}
\Delta E_{1\gamma,~2-loop~VP}(2S)=-\frac{1}{12\pi^2}\mu\alpha^2(Z\alpha)^2
\frac{W^2}{m_e^2}\int_0^1\frac{f(v)dv}{\left(1+\frac{W\sqrt{1-v^2}}{2m_e}
\right)^2}\left[1-\frac{2}{p_3(v)}+\frac{3}{2p_3^2(v)}\right]=
\end{equation}
\begin{displaymath}
=\Biggl\{{{\mu p:~-1.4112~meV}\atop{\mu d:~-1.5606~meV}},~~~p_3(v)=\frac{1}
{1+\frac{2m_e}{W\sqrt{1-v^2}}}.
\end{displaymath}
Note that, as we determine contributions to the energy spectrum numerically,
the corresponding results are given with an accuracy of 0.0001 meV.

\section{Three-loop vacuum polarization in one-photon interaction}

In the fifth order over $\alpha$ there is the contribution of three-loop
amplitudes of the vacuum polarization in the one-photon interaction (see
diagrams (a)-(b) in Fig.3). The contribution of the diagram (a) in Fig.3
to the potential takes the form:
\begin{equation}
V^C_{VP-VP-VP}(r)=-\frac{Z\alpha}{r}\frac{\alpha^3}{(3\pi)^3}\int_1^\infty
\rho(\xi)d\xi\int_1^\infty\rho(\eta d\eta\int_1^\infty\rho(\zeta)d\zeta\times
\end{equation}
\begin{displaymath}
\times\left[e^{-2m_e\zeta r}\frac{\zeta^4}{(\xi^2-\zeta^2)(\eta^2-\zeta^2)}
+e^{-2m_e\xi r}\frac{\xi^4}{(\zeta^2-\xi^2)(\eta^2-\xi^2)}+
e^{-2m_e\eta r}\frac{\eta^4}{(\xi^2-\eta^2)(\zeta^2-\eta^2)}\right].
\end{displaymath}

\begin{figure}[t!]
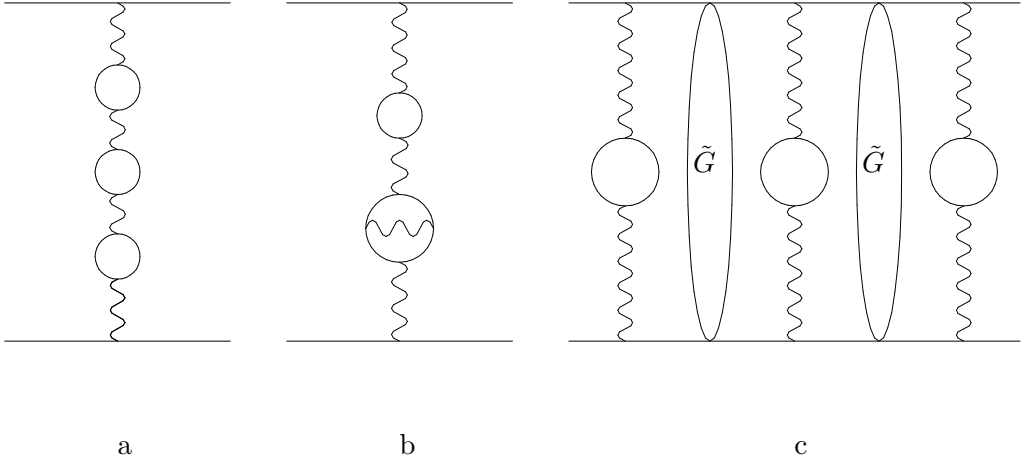

\magnitude=2000
\GRAPH(hsize=15){
\mov(0,0){\lin(2,0)}%
\mov(0,3){\lin(2,0)}%
\mov(2.5,0){\lin(2,0)}%
\mov(2.5,3){\lin(2,0)}%
\mov(5,0){\lin(4,0)}%
\mov(5,3){\lin(4,0)}%
\mov(5.5,1.5){\Circle(0.6)}%
\mov(8.5,1.5){\Circle(0.6)}%
\mov(7.,1.5){\Circle(0.6)}%
\mov(5.5,0){\wavelin(0,1.2)}%
\mov(5.5,3){\wavelin(0,-1.2)}%
\mov(8.5,0){\wavelin(0,1.2)}%
\mov(8.5,3){\wavelin(0,-1.2)}%
\mov(7.,0){\wavelin(0,1.2)}%
\mov(7.,3){\wavelin(0,-1.2)}%
\mov(6.25,1.5){\Ellipse(0.4)[1,7.5]}%
\mov(6.1,1.5){$\tilde G$}%
\mov(7.75,1.5){\Ellipse(0.4)[1,7.5]}%
\mov(7.6,1.5){$\tilde G$}%
\mov(1,1.5){\Circle(0.4)}%
\mov(1,0.75){\Circle(0.4)}%
\mov(1,2.25){\Circle(0.4)}%
\mov(3.5,1.){\Circle(0.6)}%
\mov(3.2,1.){\wavelin(0.6,0)}%
\mov(3.5,0){\wavelin(0,0.7)}%
\mov(3.5,1.3){\wavelin(0,0.5)}%
\mov(3.5,3){\wavelin(0,-0.8)}%
\mov(3.5,2.){\Circle(0.4)}%
\mov(1,0){\wavelin(0,0.55)}%
\mov(1,3){\wavelin(0,-0.55)}%
\mov(1,0){\wavelin(0,0.55)}%
\mov(1,0.95){\wavelin(0,0.35)}%
\mov(1,1.7){\wavelin(0,0.35)}%
\mov(1.,-1.){a}%
\mov(3.5,-1.){b}%
\mov(7.,-1.){c}%
}
\caption{Effects of three-loop vacuum polarization in one-photon
interaction (a,b) and in the third order of the perturbation theory (c).}
\end{figure}

The contribution of the diagram (b) in Fig.3 can be written also in the
integral form using expressions (8) and (21):
\begin{equation}
V^C_{VP-2-loop~VP}=-\frac{4\mu\alpha^3(Z\alpha)}{9\pi^3}\int_1^\infty
\rho(\xi)d\xi\int_1^\infty\frac{f(\eta)}{\eta}d\eta\frac{1}{r}
\left[e^{-2m_e\eta r}\frac{\eta^2}{\eta^2-\xi^2}-e^{-2m_e\xi r}\frac{\xi^2}
{\eta^2-\xi^2}\right].
\end{equation}
The corrections to the energy spectrum of the $(\mu p)$ and $(\mu d)$ atoms
corresponding to the interaction operators (25) and (26) are the following:
\begin{equation}
\Delta E_{VP-VP-VP}(1S)=-\frac{\mu\alpha^3(Z\alpha)^2}{27\pi^3}\int_1^\infty
\rho(\xi)d\xi\int_1^\infty\rho(\eta)d\eta\int_1^\infty\rho(\zeta)d\zeta\times
\end{equation}
\begin{displaymath}
\times\left[\frac{\xi^4}{(\xi^2-\eta^2)(\xi^2-\zeta^2)p_1^2(\xi)}+
\frac{\eta^4}{(\eta^2-\xi^2)(\eta^2-\zeta^2)p_1^2(\eta)}+
\frac{\zeta^4}{(\zeta^2-\xi^2)(\zeta^2-\eta^2)p_1^2(\zeta)}\right]=
\end{displaymath}
\begin{displaymath}
=\Biggl\{{{\mu p:~-0.0029~meV}\atop{\mu d:~-0.0034~meV}},
\end{displaymath}
\begin{equation}
\Delta E_{VP-VP-VP}(2S)=-\frac{\mu\alpha^3(Z\alpha)^2}{54\pi^3}\int_1^\infty
\rho(\xi)d\xi\int_1^\infty\rho(\eta)d\eta\int_1^\infty\rho(\zeta)d\zeta\times
\end{equation}
\begin{displaymath}
\times\Biggl\{\frac{\xi^4}{(\xi^2-\eta^2)(\xi^2-\zeta^2)}\left[\frac{1}{p_2^2(\xi)}
-\frac{1}{p_2^3(\xi)}\right]+
\frac{\eta^4}{(\eta^2-\xi^2)(\eta^2-\zeta^2)}\left[\frac{1}{p_2^2(\eta)}
-\frac{1}{p_2^3(\eta)}\right]+
\end{displaymath}
\begin{displaymath}
+\frac{\zeta^4}{(\zeta^2-\xi^2)(\zeta^2-\eta^2)}\left[\frac{1}{p_2^2(\zeta)}
-\frac{1}{p_2^3(\zeta)}\right]\Biggr\}=
\Biggl\{{{\mu p:~-0.0003~meV}\atop{\mu d:~-0.0004~meV}},
\end{displaymath}
\begin{equation}
\Delta E_{VP-2-loop~VP}(1S)=-\frac{4\mu\alpha^3(Z\alpha)^2}{9\pi^3}
\int_1^\infty\rho(\xi)d\xi\int_1^\infty\frac{f(\eta)}{\eta}\frac{d\eta}
{\eta^2-\xi^2}\left[\frac{\eta^2}{p_1^2(\eta)}-\frac{\xi^2}{p_1^2(\xi)}\right]
\end{equation}
\begin{displaymath}
=\Biggl\{{{\mu p:~-0.0223~meV}\atop{\mu d:~-0.0251~meV}},
\end{displaymath}
\begin{equation}
\Delta E_{VP-2-loop~VP}(2S)=-\frac{2\mu\alpha^3(Z\alpha)^2}{9\pi^3}
\int_1^\infty\rho(\xi)d\xi\int_1^\infty\frac{f(\eta)}{\eta}\frac{d\eta}
{\eta^2-\xi^2}\Biggl\{\eta^2\left[\frac{1}{p_2^2(\eta)}-\frac{1}{p_2^3(\eta)}
\right]-
\end{equation}
\begin{displaymath}
-\xi^2\left[\frac{1}{p_2^2(\xi)}-\frac{1}{p_2^3(\xi)}\right]\Biggr\}=
\Biggl\{{{\mu p:~-0.0030~meV}\atop{\mu d:~-0.0035~meV}}.
\end{displaymath}
There is need to make the substitution $v=\sqrt{\eta^2-1}/\eta$ in the
function $f(v)$ which is defined by Eq.(21). There exist also several
diagrams which represent three-loop corrections to the polarization operator.
They were calculated originally for the Lamb shift $(2P-2S)$ in the papers
of the Kinoshita and Nio \cite{KN1,KN2}. The greatest contribution
to the energy levels is determined by the diagrams of the sixth order
vacuum polarization with the one electron loop (the contribution $\Pi^{(p6)}$
\cite{KN1}). The estimation of their value to the $1S$ and $2S$ Lamb shift
is presented in Table I.

\section{The vacuum polarization and relativistic effects}

To calculate the energy spectra of the $S$-states of muonic hydrogen with a
precision up to terms of order $\alpha^5$ we have to construct the quasipotential
on the basis of relations (5) and (6) which catch correctly the relativistic
effects of necessary order (the Breit Hamiltonian $\Delta V_B$). Accounting
the electron vacuum polarization effects Pachucki has derived the Breit
Hamiltonian $\Delta V_B^{VP}$ in Ref. \cite{KP}. Omitting the spin-dependent
terms of the interaction operator let us write the two-body Breit Hamiltonians
$\Delta V_B$ and $\Delta V^{VP}_B$:
\begin{equation}
\Delta V_B=-{\bf p}^4\left(\frac{1}{8m_1^3}+\frac{1}{8m_2^3}\right)+\frac{\pi
Z\alpha}{2}\left(\frac{1}{m_1^2}+\frac{1}{m_2^2}\right)\delta({\bf r})-
\frac{Z\alpha}{2m_1m_2r}\left({\bf p}^2+\frac{r_ir_jp_ip_j}{r^2}\right),
\end{equation}
\begin{equation}
\Delta V_B^{VP}=\frac{\alpha}{3\pi}\int_1^\infty\rho(\xi)d\xi\Biggl\{
\frac{Z\alpha}{2}\left(\frac{1}{m_1^2}+\frac{1}{m_2^2}\right)\left[\pi\delta
({\bf r})-\frac{m_e^2\xi^2}{r}e^{-2m_e\xi r}\right]-
\end{equation}
\begin{displaymath}
-\frac{Z\alpha m_e^2\xi^2}{m_1m_2r}e^{-2m_e\xi r}(1-m_e\xi r)-
\frac{Z\alpha}{2m_1m_2}p_i\frac{e^{-2m_e\xi r}}{r}\left[\delta_{ij}+\frac{r_ir_j}
{r^2}(1+2m_e\xi r)\right]p_j.
\end{displaymath}
In the first order of the perturbation theory (FOPT) the Hamiltonian (32)
gives the following contribution after averaging over the Coulomb wave
functions (11):
\begin{equation}
\Delta E_{1,B}^{rel,VP}=<\psi_{1S,2S}|\Delta V_B^{VP}|\psi_{1S,2S}>=
\Biggl\{{{\mu p(1S):~0.1962~meV;~\mu p(2S):~0.0249~meV}\atop
{\mu d(1S):~0.2515~meV;~\mu d(2S):~0.0322~meV}}
\end{equation}
These corrections are of order $\alpha(Z\alpha)^4$. Second-order perturbation
theory (SOPT) corrections to the energy spectrum are determined by the
reduced Coulomb Green function (RCGF) \cite{VP}, whose partial expansion
is written as:
\begin{equation}
\tilde G_n({\bf r}, {\bf r'})=\sum_{l,m}\tilde g_{nl}(r,r')Y_{lm}({\bf n})
Y_{lm}^\ast({\bf n'}).
\end{equation}
The radial function $\tilde g_{nl}(r,r')$ was obtained in Ref.\cite{VP}
in the form of the Sturm expansion in the Laguerre polynomials. For the
$1S$ and $2S$-states these functions are the following:
\begin{equation}
\tilde g_{10}(r,r')=-4\mu^2 Z\alpha\left(\sum_{m=2}^\infty\frac{
L_{m-1}^1(x)L_{m-1}^1(x')}{m(m-1)}+\frac{5}{2}-\frac{x}{2}-\frac{x'}{2}\right)
e^{-\frac{x+x'}{2}},
\end{equation}
\begin{equation}
\tilde g_{20}(r,r')=-2\mu^2
Z\alpha\left[\sum_{m=1,m\not=2}^\infty
\frac{L_{m-1}^1(x)L_{m-1}^1(x')}{m(m-2)}+\left(\frac{5}{2}+x\frac{\partial}
{\partial x}+x'\frac{\partial}{\partial
x'}\right)L_1^1(x)L_1^1(x')\right]e^{-\frac{x+x'}{2}} ,
\end{equation}
where $x=\mu Z\alpha r$, $L_n^m$ are the usual Laguerre polynomials
defined as:
\begin{equation}
L_n^m(x)=\frac{e^xx^{-m}}{n!}\left(\frac{d}{dx}\right)^n\left(e^{-x}
x^{n+m}\right).
\end{equation}
As a several quasipotential terms contain $\delta({\bf r})$, we must
know $\tilde G_n({\bf r},0)$. The corresponding expression for the reduced
Coulomb Green function was obtained in Ref.\cite{KI} using the Hoestler
representation for the Coulomb Green function and subtracting the pole term.
This gave
\begin{equation}
\tilde G_{1S}({\bf r},0)=\frac{Z\alpha\mu^2}{4\pi}\frac{2e^{-x/2}}{x}
\left[2x(\ln x+C)+x^2-5x-2\right],
\end{equation}
\begin{equation}
\tilde G_{2S}({\bf r},0)=-\frac{Z\alpha\mu^2}{4\pi}\frac{e^{-x/2}}{2x}
\left[4x(x-2)(\ln x+C)+x^3-13x^2+6x+4\right],
\end{equation}
where $C=0.5772...$ is the Euler constant. One-loop vacuum polarization
gives the contribution in the second order of the perturbation theory which
is specified by the relation:
\begin{equation}
\Delta E_{n~SOPT}=2\sum_{m=1,m\not=n}^\infty\frac{<\psi_n^C|\Delta V_B|
\psi_m^C><\psi_m^C|V_{VP}^C|\psi_n^C>}{E_n^C-E_m^C},
\end{equation}
The matrix element of the operator ${\bf p}^4$ is expressed by means of the
substitution ${\bf p}^4/4\mu^2$= $(H_0+Z\alpha/r)(H_0+Z\alpha/r)$
($H_0={\bf p}^2/2\mu$ - $Z\alpha/r$). After that we employ the sequence of
algebraic transformations:
\begin{equation}
<\psi_n^C|\frac{Z\alpha}{r}H_0\sum_{m,m\not =n}\frac{|\psi_m^C><\psi_m^C|}
{E_n^C-E_m^C}V^c_{VP}|\psi_n^C>=
\end{equation}
\begin{displaymath}
=-<\psi_n^C|\frac{Z\alpha}{r}\left(I-|\psi_n^C><\psi_n^C|\right)V^C_{VP}|\psi_n^C>+
E_n^C<\psi_n^C|\frac{Z\alpha}{r}\sum_{m,m\not =n}\frac{|\psi_m^C><\psi_m^C|}
{E_n^C-E_m^C}V^C_{VP}|\psi_n^C>.
\end{displaymath}
Integrating the RCGF (34) over the coordinates $r$ and $r'$ we obtain the
sums which are calculated accounting the range of a variable $\xi$
$(1\div\infty)$ and $1/(1+W/m_e\xi)<1$. Typical matrix element for the $1S$
correction is the following:
\begin{equation}
I=\int_0^\infty\left(1-\frac{x}{2}\right)e^{-x}dx\int_0^\infty x' e^{-x'p_1(\xi)}
dx'\left[\frac{5}{2}-\frac{x}{2}-\frac{x'}{2}+\sum\frac{L_{m-1}^1(x)L_{m-1}^1(x')}
{m(m-1)}\right]=
\end{equation}
\begin{displaymath}
=\frac{5}{4p_1^2(\xi)}-\frac{1}{2p_1^3(\xi)}+\frac{\ln p_1(\xi)}{p_1^2(\xi)},
\end{displaymath}
\begin{displaymath}
\sum_{m=2}^\infty\frac{[p_1(\xi)-1]^{m-1}}{(m-1)p_1(\xi)^{m+1}}=\frac{1}{p_1^2(\xi)}
\sum_{m=2}^\infty\frac{1}{(m-1)p_1^{m-1}(\xi)}=\frac{\ln p_1(\xi)}{p_1^2(\xi)}.
\end{displaymath}
Omitting other numerous intermediate analytical expressions we write the
summary numerical value of the second order of the perturbation theory
contribution (40) for the $1S$ and $2S$-levels:
\begin{equation}
\Delta
E_{2,'}^{rel,VP}=\Biggl\{{{\mu p(1S):~-0.2644~meV;~\mu p(2S):~-0.0559~meV}\atop
{\mu d(1S):~-0.3194~meV;~\mu d(2S):~-0.0696~meV}}
\end{equation}

\section{Two-loop and three-loop vacuum polarization in the second order
of the perturbation theory}

Two-loop vacuum polarization gives the correction in the second order
of the perturbation theory (see the diagram (a) in Fig.4):
\begin{equation}
\Delta E^{VP,VP}_{SOPT}=<\psi_n^C|V^C_{VP}\cdot \tilde G\cdot V^C_{VP}|\psi^C_n>.
\end{equation}
The calculation of this matrix element can be done with the relations
(10), (35) and (36). As a result we obtain the corrections of order
$\alpha^2(Z\alpha)^2$ for the $1S$ and $2S$ states:
\begin{equation}
\Delta E^{VP,VP}_{SOPT}(1S)=-\frac{\mu\alpha^2(Z\alpha)^2}{9\pi^2}
\int_1^\infty\rho(\xi)d\xi\int_1^\infty\rho(\eta)d\eta
\Biggl[\frac{5}{2}\frac{1}{p_1^2(\xi)p_1^2(\eta)}-
\end{equation}
\begin{displaymath}
-\frac{1}{p_1^2(\xi)
p_1^3(\eta)}-\frac{1}{p_1^3(\xi)p_1^2(\eta)}+\frac{[p_1(\xi)-1][p_1(\eta)-1]}
{p_1^2(\xi)p_1^2(\eta)(p_1(\xi)+p_1(\eta)-1)}-\frac{1}{p_1^2(\xi)p_1^2(\eta)}
\ln\frac{[p_1(\xi)+p_1(\eta)-1]}{(p_1(\xi)+p_1(\eta))}\Biggr],
\end{displaymath}
\begin{equation}
\Delta E^{VP,VP}_{SOPT}(2S)=-\frac{\mu\alpha^2(Z\alpha)^2}{9\pi^2}
\int_1^\infty\rho(\xi)d\xi\int_1^\infty\rho(\eta)d\eta
f(\xi,\eta),
\end{equation}
\begin{displaymath}
f(\xi,\eta)=-\frac{9}{2p_2^4(\xi)p_2^5(\eta)}+\frac{6}{p_2^3(\xi)p_2^5(\eta)}-
\frac{3}{p_2^2(\xi)p_2^5(\eta)}-\frac{9}{2p_2^5(\xi)p_2^4(\eta)}+\frac{153}
{8p_2^4(\xi)p_2^4(\eta)}-\frac{39}{2p_2^3(\xi)p_2^4(\eta)}+
\end{displaymath}
\begin{displaymath}
+\frac{33}{4p_2^2(\xi)p_2^4(\eta)}
+\frac{6}{p_2^5(\xi)p_2^3(\eta)}-\frac{39}{2p_2^4(\xi)p_2^3(\eta)}+
\frac{17}{p_2^3(\xi)p_2^3(\eta)}-\frac{6}{p_2^2(\xi)p_2^3(\eta)}-
\frac{3}{p_2^5(\xi)p_2^2(\eta)}+
\end{displaymath}
\begin{displaymath}
+\frac{33}{4p_2^4(\xi)p_2^2(\eta)}-
\frac{6}{p_2^3(\xi)p_2^2(\eta)}+\frac{3}{2p_2^2(\xi)p_2^2(\eta)}
-\frac{z^4}{a_2^2(\xi)a_2^2(\eta)(z-1)}-\frac{z^4(6-5z)}
{2a_2^2(\xi)a_2^2(\eta)(z-1)^2p_2(\xi)}-
\end{displaymath}
\begin{displaymath}
-\frac{z^4(6-5z)}{2a_2^2(\xi)a_2^2(\eta)(z-1)^2p_2(\eta)}
+\frac{z^4(4-3z)}{2a_2^2(\xi)a_2^3(\eta)(z-1)^2}+\frac{z^4(4-3z)}
{2a_2^3(\xi)a_2^2(\eta)(z-1)^2}+\frac{z^5(-30+49z-21z^2)}
{4a_2^3(\xi)a_2^3(\eta)(z-1)^3}-
\end{displaymath}
\begin{displaymath}
-\frac{z^4(-18+27z-11z^2)}{4a_2^2(\xi)a_2^3(\eta)(z-1)^3p_2(\xi)}
-\frac{z^4(-18+27z-11z^2)}{4a_2^3(\xi)a_2^2(\eta)(z-1)^3p_2(\eta)}+
\frac{z^4(-10+13z-5z^2)}{4a_2^3(\xi)a_2^3(\eta)(z-1)^3}
+\ln(1-z)\times
\end{displaymath}
\begin{displaymath}
\times\Biggl[-\frac{2z^3}{a_2^2(\xi)a_2^2(\eta)}
+\frac{3z^3}{a_2^2(\xi)a_2^2(\eta)p_2(\xi)}+
\frac{3z^3}{a_2^2(\xi)a_2^2(\eta)p_2(\eta)}
-\frac{z^3}{a_2^2(\xi)a_2^3(\eta)}-\frac{z^3}{a_2^3(\xi)a_2^2(\eta)}-
\end{displaymath}
\begin{displaymath}
\frac{9z^4}{2a_2^3(\xi)a_2^3(\eta)}+\frac{3z^3}{2a_2^2(\xi)a_2^3(\eta)p_2(\xi)}+
\frac{3z^3}{2a_2^3(\xi)a_2^2(\eta)p_2(\eta)}-\frac{z^3}{2a_2^3(\xi)a_2^3(\eta)}
\Biggr],
\end{displaymath}
where $a_2(\xi)=p_2(\xi)-1$, $z=a_2(\xi)a_2(\eta)/p_2(\xi)p_2(\eta)$.

\begin{figure}[t!]
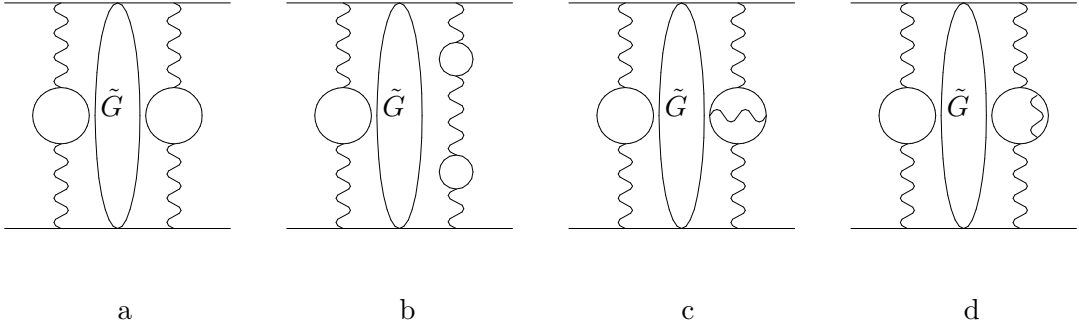

\magnitude=2000
\GRAPH(hsize=15){
\mov(0,0){\lin(2,0)}%
\mov(0,2){\lin(2,0)}%
\mov(2.5,0){\lin(2,0)}%
\mov(2.5,2){\lin(2,0)}%
\mov(5,0){\lin(2,0)}%
\mov(5,2){\lin(2,0)}%
\mov(3.,1.){\Circle(0.5)}%
\mov(3.,0){\wavelin(0.,0.75)}%
\mov(3.,2.){\wavelin(0.,-0.75)}%
\mov(4.,0.5){\Circle(0.3)}%
\mov(4.,1.5){\Circle(0.3)}%
\mov(4.,0){\wavelin(0.,0.35)}%
\mov(4.,2){\wavelin(0.,-0.35)}%
\mov(4.,0.65){\wavelin(0.,0.7)}%
\mov(7.5,0){\lin(2,0)}%
\mov(7.5,2){\lin(2,0)}%
\mov(6,1){\Ellipse(0.4)[1,5]}%
\mov(5.85,1){${\tilde G}$}%
\mov(0.85,1.){${\tilde G}$}%
\mov(1.,1){\Ellipse(0.4)[1,5]}%
\mov(3.5,1){\Ellipse(0.4)[1,5]}%
\mov(3.35,1.){${\tilde G}$}%
\mov(1.5,1.){\Circle(0.5)}%
\mov(8.5,1){\Ellipse(0.4)[1,5]}%
\mov(8.35,1){${\tilde G}$}%
\mov(1.,-0.8){a}%
\mov(3.5,-0.8){b}%
\mov(6.,-0.8){c}%
\mov(8.5,-0.8){d}%
\mov(1.5,0){\wavelin(0,0.75)}%
\mov(1.5,2){\wavelin(0,-0.75)}%
\mov(6.5,0){\wavelin(0.,0.75)}%
\mov(6.5,2){\wavelin(0.,-0.75)}%
\mov(6.5,1.){\Circle(0.5)}%
\mov(6.25,1){\magnitude=1500\wavelin(0.5,0)}%
\mov(9,0){\wavelin(0,0.75)}%
\mov(9,2){\wavelin(0,-0.75)}%
\mov(9.,1.){\Circle(0.5)}%
\mov(9.15,0.81){\magnitude=1500\wavelin(0.,0.37)}%
\mov(0.5,1.){\Circle(0.5)}%
\mov(0.5,0){\wavelin(0,0.75)}%
\mov(0.5,1.25){\wavelin(0,0.75)}%
\mov(5.5,0){\wavelin(0,0.75)}%
\mov(5.5,1.25){\wavelin(0,0.75)}%
\mov(5.5,1.){\Circle(0.5)}%
\mov(8,0){\wavelin(0,0.75)}%
\mov(8,1.25){\wavelin(0,0.75)}%
\mov(8,1.){\Circle(0.5)}%
}
\caption{Corrections of two-loop and three-loop vacuum polarization
in the second order of the perturbation theory.}
\end{figure}

Three-loop vacuum polarization contributions in the second order of the
perturbation theory shown in Fig.4(b,c,d) are of order $\alpha^3(Z\alpha)^2$.
For their calculation we must use again the analytical expression for the
one-loop and two-loop polarization operator. Fulfilling the integration
over the wave function coordinates we represent the necessary corrections
in the form:
\begin{equation}
\Delta E_{SOPT}^{VP-VP,VP}(1S)=-\frac{2\mu\alpha^3(Z\alpha)^2}{27\pi^3}\int_1^\infty
\rho(\xi)d\xi\int_1^\infty\frac{\rho(\eta)d\eta}{(\xi^2-\eta^2)}
\int_1^\infty\frac{\rho(\zeta)d\zeta}
{p_1^2(\zeta)}[\frac{\xi^2g(\xi,\zeta)}{p_1^2(\xi)}-
\frac{\eta^2 g(\eta,\zeta)}{p_1^2(\eta)}],
\end{equation}
\begin{equation}
g(\xi,\zeta)=\frac{5}{2}-\frac{1}{p_1(\xi)}-\frac{1}{p_1(\zeta)}+
\frac{[p_1(\xi)-1][p_1(\zeta)-1]}{p_1(\xi)+p_1(\zeta)-1}-\ln\frac{p_1(\xi)+
p_1(\zeta)-1}{p_1(\xi)\cdot p_1(\zeta)},
\end{equation}
\begin{equation}
\Delta E_{SOPT}^{VP-VP,VP}(2S)=-\frac{2\mu\alpha^3(Z\alpha)^2}{27\pi^3}\int_1^\infty
\rho(\xi)d\xi\int_1^\infty\rho(\eta)d\eta\int_1^\infty\rho(\zeta)d\zeta
\frac{[\xi^2f(\xi,\zeta)-\eta^2 f(\eta,\zeta)]}
{\xi^2-\eta^2},
\end{equation}
where the function $f(\xi,\zeta)$ is defined earlier by the Eq.(46),
\begin{equation}
\Delta E_{SOPT}^{2-loop~VP,VP}(1S)=-\frac{\mu\alpha^3(Z\alpha)^2}{9\pi^2}\int_0^1f(v)dv
\Biggl\{\frac{5}{2}\frac{1}{p_1^2(\xi)q_1^2(v)}-\frac{1}{p_1^3(\xi)q_1^2(v)}+
\end{equation}
\begin{displaymath}
+\frac{1}{p_1^2(\xi)q_1^2(v)}\left[\frac{p_1(\xi)q_1(v)}{p_1(\xi)+q_1(v)+1}-
\ln\frac{p_1(\xi)+q_1(v)+1}{p_1(\xi)+q_1(v)}\right]\Biggr\},~q_1(v)=1+\frac{m_e}
{W\sqrt{1-v^2}},
\end{displaymath}
\begin{equation}
\Delta E_{SOPT}^{2-loop~VP,VP}(2S)=-\frac{4\mu\alpha^3(Z\alpha)^2}{9\pi^3}\int_0^1\frac{f(v)dv}
{1-v^2}\int_1^\infty\rho(\xi)d\xi f(\xi,v),
\end{equation}
where $p_2(v)=1+2m_e/W\sqrt{1-v^2}$ in the function $f(\xi,v)$,
$a_2(v)=2m_e/W\sqrt{1-v^2}$. The energy terms (45)-(51) give the
contribution to the $(\mu p)$ - $(\mu d)$ isotope shift:
\begin{equation}
\Delta E_{SOPT,IS}^{VP,VP}=0.3125~meV,
\end{equation}
and the individual values for the $S$-states are depicted in Table I.
The analysis of the three-loop vacuum polarization contribution in the
third order of the perturbation theory (see the diagram (c) in Fig.3)
shows that it is one order of the magnitude smaller than the contribution
of the amplitude shown in Fig.4(c) and can be neglected.

\section{Muon self-energy, muon vacuum polarization
and recoil corrections}

To estimate the corrections of the muon self-energy (MSE) and muon vacuum
polarization (MVP) of order $\alpha(Z\alpha)^4$ we used well-known analytical
results for the hydrogen atom. In the case of the $1S$ and $2S$-states
these contributions are the following \cite{EGS}:
\begin{equation}
\Delta E_{MVP,MSE}(1S)=\frac{\alpha(Z\alpha)^4}{\pi}\frac{\mu^3}{m_1^2}
\left[\frac{4}{3}\ln\frac{m_1}{\mu(Z\alpha)^2}-\frac{4}{3}\ln k_0(1,0)+
\frac{38}{45}\right]=
\Biggl\{{{\mu p:~5.1180~meV}\atop{\mu d:~5.9395~meV}},
\end{equation}
\begin{equation}
\Delta E_{MVP,MSE}(2S)=\frac{\alpha(Z\alpha)^4}{8\pi}\frac{\mu^3}{m_1^2}
\left[\frac{4}{3}\ln\frac{m_1}{\mu(Z\alpha)^2}-\frac{4}{3}\ln k_0(2,0)+
\frac{38}{45}\right]=
\Biggl\{{{\mu p:~0.6543~meV}\atop{\mu d:~0.7594~meV}},
\end{equation}
where $\ln k_0(n,l)$ is the Bethe logarithm:
\begin{equation}
\ln k_0(1,0)=2.984128555765498,~~~\ln k_0(2,0)=2.811769893120563.
\end{equation}
The correction determined as a sum of radiative insertions in the lepton line
is known also in analytical form. We represent its value despite the fact that
the order of this correction is $\alpha(Z\alpha)^5$ because the corresponding
coefficient is sufficiently large (of order 10) \cite{EGS}:
\begin{equation}
\Delta E_{rad}(1S)=\frac{\alpha(Z\alpha)^5\mu^3}{m_1^2}\left(\frac{427}{96}-
2\ln 2\right)=\Biggl\{{{\mu p:~0.0355~meV}\atop{\mu d:~0.0414~meV}}
\end{equation}
\begin{equation}
\Delta E_{rad}(2S)=\frac{\alpha(Z\alpha)^5\mu^3}{8m_1^2}\left(\frac{427}{96}-
2\ln 2\right)=\Biggl\{{{\mu p:~0.0044~meV}\atop{\mu d:~0.0052~meV}}.
\end{equation}
Following the Pachucki paper \cite{KP}, let us estimate also the contributions
of two diagrams of the sixth order over $\alpha$ enhanced by the $\ln\alpha$.
The diagram (a) in Fig.5 of the muon radiative correction with electron
vacuum polarization gives the following contributions:
\begin{equation}
\Delta E_{1,rad+VP}(1S)=\frac{2\mu^3\alpha^2(Z\alpha)^4}{9\pi^2m_1^2}
\ln\frac{m_1}{\mu\alpha^2}\int_1^\infty\rho(\xi)d\xi\left[\frac{2}{p_1(\xi)}-
\frac{2}{p_1^2(\xi)}
+\frac{1}{p_1^3(\xi)}\right]=
\end{equation}
\begin{displaymath}
=\Biggl\{{{\mu p:~0.0061~meV}\atop{\mu d:~0.0073~meV}},
\end{displaymath}
\begin{equation}
\Delta E_{1,rad+VP}(2S)=\frac{\mu^3\alpha^2(Z\alpha)^4}{48\pi^2m_1^2}
\ln\frac{m_1}{\mu\alpha^2}\int_1^\infty\rho(\xi)d\xi\times
\end{equation}
\begin{displaymath}
\times\left[\frac{16}{3p_2(\xi)}-\frac{9}{2p_2^2(\xi)}
+\frac{3}{p_2^3(\xi)}-\frac{9}{4p_2^4(\xi)}+\frac{1}{p_2^5(\xi)}\right]=
\Biggl\{{{\mu p:~0.0010~meV}\atop{\mu d:~0.0012~meV}}
\end{displaymath}

\begin{figure}[t!]
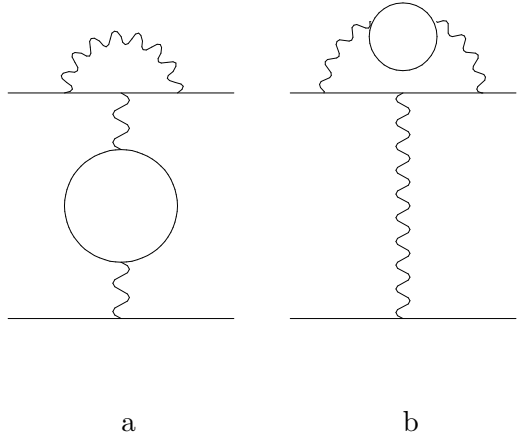

\magnitude=2000
\GRAPH(hsize=15){
\mov(3,1){\lin(2,0)}%
\mov(3,3){\lin(2,0)}%
\mov(5.5,1){\lin(2,0)}%
\mov(5.5,3){\lin(2,0)}%
\mov(4.,2.){\Circle(1.)}%
\mov(4.,1){\wavelin(0,0.5)}%
\mov(4.,2.5){\wavelin(0,0.5)}%
\mov(6.5,1){\wavelin(0,2)}%
\mov(4.,3.){\halfwavecirc(1.)[+]}%
\mov(6.5,3.){\wavearc(0.7)[115,180]}%
\mov(6.5,3.){\wavearc(0.7)[0,65]}%
\mov(6.5,3.5){\Circle(0.6)}%
\mov(4.,-0.){a}%
\mov(6.5,-0.){b}%
}
\caption{Radiative corrections accounting electron vacuum polarization.}
\end{figure}

The diagram (b) in Fig.5 gives the contribution to the energy spectrum
which can be expressed in terms of the slope of the Dirac form factor
$F_1'$ and the Pauli form factor $F_2$:
\begin{equation}
\Delta E_{2,rad+VP}(nS)=\frac{\mu^3}{m_1^2}\frac{(Z\alpha)^4}{n^3}\left[4m_1^2
F_1'(0)+F_2(0)\right],
\end{equation}
where the form factors $F_1'$ and $F_2$ were calculated in Ref.\cite{BCR}:
\begin{equation}
m_1^2F_1'(0)=\left(\frac{\alpha}{\pi}\right)^2\left[\frac{1}{9}\ln^2\frac
{m_1}{m_e}-\frac{29}{108}\ln\frac{m_1}{m_e}+\frac{1}{9}\zeta(2)+\frac{395}
{1296}\right],
\end{equation}
\begin{equation}
F_2(0)=\left(\frac{\alpha}{\pi}\right)^2\left[\frac{1}{3}\ln\frac{m_1}{m_e}
-\frac{25}{36}\right].
\end{equation}
Then the correction to the Lamb shift of the S-levels takes the form:
\begin{equation}
\Delta E_{2,rad+VP}(nS)=\frac{\mu^3\alpha^2(Z\alpha)^4}{m_1^2n^3\pi^2}\left(
\frac{4}{9}\ln^2\frac{m_1}{m_e}-\frac{20}{27}\ln\frac{m_1}{m_e}+\frac{4}{9}
\zeta(2)+\frac{85}{162}\right).
\end{equation}

In this section we included yet known analytical recoil correction of order
$(Z\alpha)^5$ which is determined by two-photon exchange diagrams \cite{SE}:
\begin{equation}
\Delta E_{rec}(nS)=\frac{(Z\alpha)^5\mu^3}{\pi n^3m_1m_2}\Biggl[-\frac{2}{3}\ln(Z\alpha)-
\frac{8}{3}\ln k_0(n,0)-\frac{1}{9}-\frac{7}{3}a_n-
\end{equation}
\begin{displaymath}
-\frac{2}{m_2^2-m_1^2}\left(
m_2^2\ln\frac{m_1}{\mu}-m_1^2\ln\frac{m_2}{\mu}\right)\Biggr],
\end{displaymath}
where
\begin{equation}
a_n=-2\left[\ln\frac{2}{n}+\left(1+\frac{1}{2}+...+\frac{1}{n}\right)+1-\frac{1}{2n}\right].
\end{equation}
Numerical values of the corrections (63) and (64) are presented in Table I.

\section{Effects of nuclear structure, polarizability and vacuum polarization}

In the energy spectrum of muonic hydrogen the important role belongs to
strong interactions which are connected with the distributions of electric charge
and magnetic moment of the nucleus. Namely these corrections together with
the mass differences of the isotopes lead to the isotope shift in the system
$(\mu p)$ - $(\mu d)$. In the leading order $(Z\alpha)^4$ the nuclear structure
effects are determined by the differential parameter of the electric charge
distribution known as the nucleus charge radius $r_N$. To calculate one-loop
corrections where the nucleus structure effects are essential we must know
the form of the nucleus electromagnetic form factors. The contribution of
the nuclear structure effects  was studied both for the hyperfine structure
and the Lamb shift in hydrogenic atoms in Refs.\cite{KP,FM,FM3,Friar,EGS}.
The leading order nuclear structure correction of order $(Z\alpha)^4$ to the
Lamb shift of the $S$-levels in muonic hydrogen has the form (see the diagram
(a) in Fig.6):
\begin{equation}
\Delta E_{str,(Z\alpha)^4}(nS)=\frac{2}{3n^3}\mu^3(Z\alpha)^4<r_p^2>,
\end{equation}
where $r_p^2$ is the mean-square proton charge radius. Numerical values
of the correction (66) for the levels with $n=1$ and $n=2$ (at $r_p$=0.891 fm
\cite{EGS}) are presented in Table I. They give essential relative order
contribution, so, the decrease of the error in the value of the proton charge
radius is extremely important task if we desire to obtain more precise
theoretical values of the Lamb shift of the $S$-states, different fine
structure intervals and the isotope shifts. In the case of the muonic
deuterium the deuteron charge radius $r_d$=2.094 fm is used \cite{FM3}.

\begin{figure}[t!]
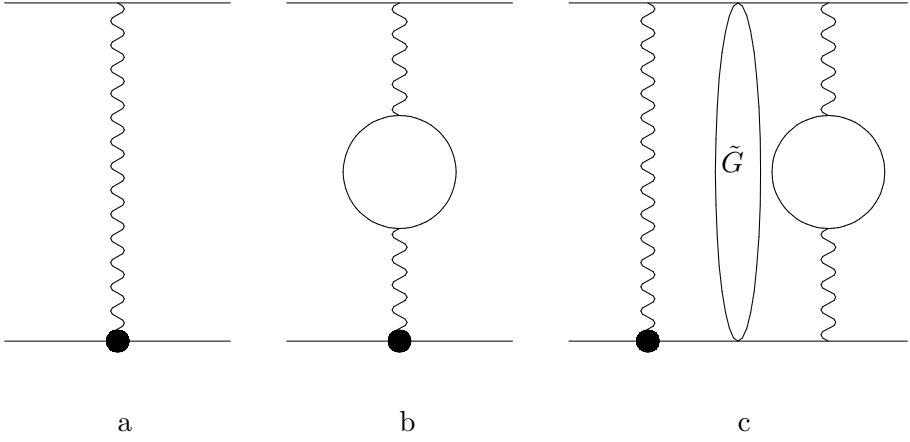

\magnitude=2000
\GRAPH(hsize=15){
\mov(0,0){\lin(2,0)}%
\mov(0,3){\lin(2,0)}%
\mov(2.5,0){\lin(2,0)}%
\mov(2.5,3){\lin(2,0)}%
\mov(5,0){\lin(3,0)}%
\mov(5,3){\lin(3,0)}%
\mov(1.,0.){\Circle**(0.2)}%
\mov(3.5,0){\Circle**(0.2)}%
\mov(5.7,0.){\Circle**(0.2)}%
\mov(1.,-0.8){a}%
\mov(3.5,-0.8){b}%
\mov(6.5,-0.8){c}%
\mov(1,0){\wavelin(0,3)}%
\mov(3.5,0){\wavelin(0,1)}%
\mov(3.5,2){\wavelin(0,1)}%
\mov(5.7,0){\wavelin(0,3)}%
\mov(7.3,0){\wavelin(0,1)}%
\mov(7.3,2){\wavelin(0,1)}%
\mov(6.5,1.5){\Ellipse(0.4)[1,7.5]}%
\mov(6.35,1.5){${\tilde G}$}%
\mov(3.5,1.5){\Circle(1.)}%
\mov(7.3,1.5){\Circle(1.)}%
}
\caption{The nuclear structure and vacuum polarization corrections.}
\end{figure}

Two-photon exchange amplitudes shown in Fig.7 give the nuclear structure
corrections of order $(Z\alpha)^5$. In this case the quasipotential of two-photon
interaction can be found by means of the relations (5) and (6) \cite{FM,KP}.
Corresponding correction to the energy spectrum takes the form of one-dimensional
integral:
\begin{equation}
\Delta E_{str,(Z\alpha)^5}(nS)=-\frac{\mu^3(Z\alpha)^5}{\pi n^3}\int_0^\infty
\frac{dk}{k}V(k),
\end{equation}
\begin{equation}
V(k)=\frac{2(F_1^2-1)}{m_1m_2}+\frac{8m_1[F_2(0)+4m_2^2F_1'(0)]}{m_2(m_1+m_2)k}
+\frac{k^2}{2m_1^3m_2^3}\times
\end{equation}
\begin{displaymath}
\times\left[2(F_1^2-1)(m_1^2+m_2^2)+4F_1F_2m_1^2+3F_2^2m_1^2\right]
+\frac{\sqrt{k^2+4m_1^2}}{2m_1^3m_2(m_1^2-m_2^2)k}\times
\end{displaymath}
\begin{displaymath}
\times\Biggl\{k^2\left[2(F_1^2-1)m_2^2+4F_1F_2m_1^2+3F_2^2m_1^2\right]
-8m_1^4F_1F_2+\frac{16m_1^4m_2^2(F_1^2-1)}{k^2}\Biggr\}-
\end{displaymath}
\begin{displaymath}
-\frac{\sqrt{k^2+4m_2^2}m_1}{2m_2^3(m_1^2-m_2^2)k}\Biggl\{k^2\left[2(F_1^2-1)+
4F_1F_2+3F_2^2\right]
-8m_2^2F_1F_2+\frac{16m_2^4(F_1^2-1)}{k^2}\Biggr\}.
\end{displaymath}
To carry out numerical integration in Eq.(67) we used the parameterization
of the Dirac $F_1$ and Pauli $F_2$ proton form factors obtained in Ref.\cite{Simon}.
For the muonic deuterium the similar contributions are obtained in Ref.\cite{FM3}.
The fifth order over $\alpha$ contribution connected with the nucleus structure
is given also by the effects of the electron vacuum polarization which are
shown in Fig.6(b,c). The particle interaction operator corresponding the
amplitude in Fig.6(b) is determined by the following expression:
\begin{equation}
\Delta V_{str,VP}(r)=\frac{2\alpha(Z\alpha)r_N^2}{9}\int_1^\infty\rho(\xi)
d\xi\left[\pi\delta({\bf r})-\frac{m_e^2\xi^2}{r}e^{-2m_e\xi r}\right].
\end{equation}
The matrix elements of the operator (69) over the functions (11) lead
to numerical data:
\begin{equation}
\Delta E_{str,VP}(1S)=\frac{2}{9}\alpha(Z\alpha)^4\mu^3 r_N^2\int_1^\infty
\frac{\rho(\xi)d\xi}{p_1^2(\xi)}(1+\frac{2m_e\xi}{W})=
\Biggl\{{{\mu p:~0.1991~meV}\atop{\mu d:~1.4155~meV}},
\end{equation}
\begin{equation}
\Delta E_{str,VP}(2S)=\frac{1}{36}\alpha(Z\alpha)^4\mu^3 r_N^2\int_1^\infty
\frac{\rho(\xi)d\xi}{p_1^2(\xi)}\left[(1+\frac{4m_e\xi}{W})-\frac{4m_e^2\xi^2}
{W^2}\left(-\frac{2}{p_2(\xi)}+\frac{3}{2p_2^2(\xi)}\right)\right]=
\end{equation}
\begin{displaymath}
=\Biggl\{{{\mu p:~0.0257~meV}\atop{\mu d:~0.1824~meV}}.
\end{displaymath}
There exist also the contribution of the electron vacuum polarization
and nuclear structure in the second order of the perturbation theory
(see the diagram (c) in Fig.6) which is determined by the reduced Coulomb
Green function $\tilde G_n(r,0)$ (38) and (39). In this case the contributions
to the shifts of the $S$-levels are:
\begin{equation}
\Delta E_{str,VP;SOPT}(1S)=-\frac{2}{9\pi}\alpha(Z\alpha)^4\mu^3 r_N^2
\int_1^\infty\frac{\rho(\xi)d\xi}{p_1^3(\xi)}\Bigl[2-3p_1(\xi)-2p_1^2(\xi)-2p_1(\xi)\ln
p_1(\xi)\Bigr]=
\end{equation}
\begin{displaymath}
=\Biggl\{{{\mu p:~0.1242~meV}\atop{\mu d:~0.8913~meV}},
\end{displaymath}
\begin{equation}
\Delta E_{str,VP;SOPT}(2S)=\frac{1}{36\pi}\alpha(Z\alpha)^4\mu^3 r_N^2
\int_1^\infty\frac{\rho(\xi)d\xi}{p_1^5(\xi)}\Bigl[-12+23p_2(\xi)-8p_2^2(\xi)-
4p_2^3(\xi)+4p_2^4(\xi)+
\end{equation}
\begin{displaymath}
+4p_2(\xi)(3-4p_2(\xi)+2p_2^2(\xi))\ln p_2(\xi)\Bigr]=
\Biggl\{{{\mu p:~0.0126~meV}\atop{\mu d:~0.0898~meV}}.
\end{displaymath}

\begin{figure}[t!]
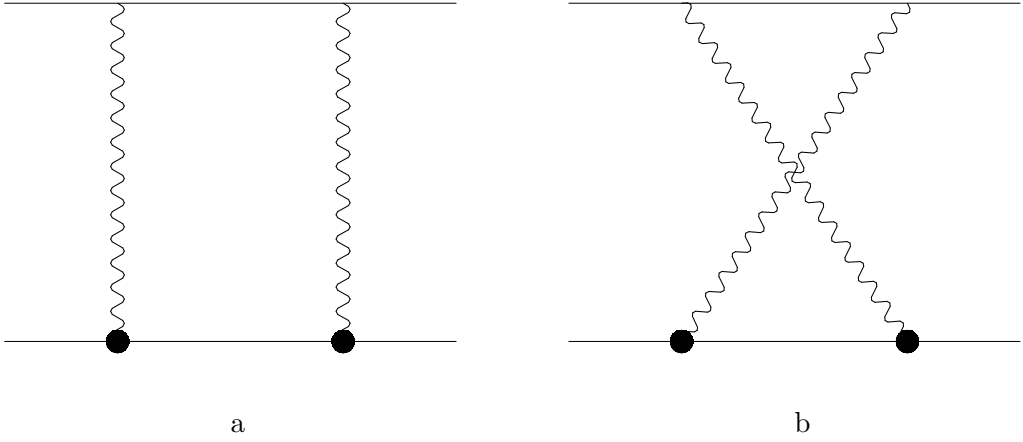

\magnitude=2000
\GRAPH(hsize=15){
\mov(0,0){\lin(1,0)}%
\mov(3,0){\lin(1,0)}%
\mov(0,3){\lin(4,0)}%
\mov(5,0){\lin(1,0)}%
\mov(8,0){\lin(1,0)}%
\mov(5,3){\lin(4,0)}%
\mov(3.,0.){\Circle**(0.2)}%
\mov(6.,0){\Circle**(0.2)}%
\mov(1.,0.){\Circle**(0.2)}%
\mov(8.,0.){\Circle**(0.2)}%
\mov(2.,-0.8){a}%
\mov(7.,-0.8){b}%
\mov(1.,0){\lin(2.,0)}%
\mov(6.,0){\lin(2.,0)}%
\mov(1,0){\wavelin(0,3)}%
\mov(3,0){\wavelin(0,3)}%
\mov(6,0){\wavelin(2,3)}%
\mov(8,0){\wavelin(-2,3)}%
}
\caption{The nuclear structure corrections of order $(Z\alpha)^5$.
Bold circle on the diagram denotes the nucleus vertex operator.}
\end{figure}

The strong interaction contribution to the energy spectrum of the $(\mu p)$
and $(\mu d)$ comes not only from the nucleus structure but also from
the nucleus polarizability. For muonic deuterium atom we used analytical
expression of the deuteron polarizability correction obtained in Ref.\cite{MPK}:
\begin{equation}
\Delta E_{d,pol}(nS)=-\frac{\alpha(Z\alpha)^3}{\pi n^3}\mu^3m_1\left[
5\alpha_d\left(\ln\frac{8I}{m_1}+\frac{1}{20}\right)-\beta_d\left(\ln\frac
{8I}{m_1}-1.24\right)\right],
\end{equation}
where the deuteron bound energy $I=\kappa^2/m_2$, and $\kappa=45.7$ MeV,
$\alpha_d=0.635~fm^3$ and $\beta_d=0.073~fm^3$ are the electric and magnetic
deuteron polarizabilities. For muonic hydrogen atom the proton polarizability
contribution to the Lamb shift of the $S$-states can be calculated employing
the expression \cite{FM,KP1999}:
\begin{equation}
\Delta E_{p,pol}(nS)=-\frac{16\mu^3(Z\alpha)^5m_1}{\pi^2n^3}\int_0^\infty
\frac{dk}{k}\int_0^{\pi}d\phi\int_{\nu_0}^\infty dy\frac{\sin^2\phi}
{(k^2+4m_1^2\cos^2\phi)(y^2+k^2\cos^2\phi)}\times
\end{equation}
\begin{displaymath}
\times[(1+2\cos^2\phi)\frac{(1+k^2/y^2)\cos^2\phi}{1+R(y,k^2)}+
\sin^2\phi]F_2(y,k^2)+\frac{2\mu^3\alpha^5}{\pi n^3m_1m_2}\int_0^\infty
h(k^2)\beta(k^2) k dk,
\end{displaymath}
where $F_2$ is the structure function of the lepton-nucleon scattering,
$R=\sigma_L/\sigma_T$ is the ratio of the cross sections for the absorption
of longitudinally and transversely polarized photons by hadrons,
$\nu_0$ is the threshold for the $\pi$-meson production. The subtraction
term in Eq.(75) includes the function
\begin{equation}
h(k^2)=1+\left(1-\frac{k^2}{2m_1^2}\right)\left(\sqrt{\frac{4m_1^2}
{k^2}+1}-1\right), \beta(k^2)=\beta \cdot G(k^2), G(k^2)=\frac{1}
{\left(1+k^2/0.71\right)^2},
\end{equation}
where $\beta_p=2.1(0.9)\times 10^{-4}~fm^3$ is the proton magnetic polarizability.
The contribution of internal deuteron polarizability can be calculated
also on the basis of Eq.(75) \cite{FM4}. It is included in total value together
with the correction (74) which is dominating.

Hadronic vacuum polarization contribution of order $\alpha(Z\alpha)^4$
to the shift of $S$-levels in muonic hydrogen which was studied in
Ref.\cite{FMS,FM2} can be presented in the form:
\begin{equation}
\Delta E_{HVP}(nS)=-\frac{4\alpha(Z\alpha)^4\mu^3}{\pi n^3}\int_{4m_\pi^2}
^\infty\frac{\rho^h(s)ds}{s},
\end{equation}
where the spectral function $\rho^h(s)$ is expressed through the cross
section of $e^+e^-$ annihilation into hadrons:
\begin{equation}
\rho^h(s)=\frac{\sigma^h(e^+e^-\to hadrons)}{3s\sigma_{\mu\mu}(e^+e^-\to
\mu^+\mu^-)}.
\end{equation}
Dividing the whole integration region over $s$ into the intervals where the
cross section of $e^+e^-$-annihilation into hadrons is known from the
experiment \cite{CMD} we can carry out numerical integration in Eq.(77).
Total numerical results are included in Table I.

\section{Conclusion}

\begin{table}[!ht]
\caption{\label{t1}Corrections of order $\alpha^3$, $\alpha^4$ and $\alpha^5$
to the $1S$ and $2S$ Lamb shifts in muonic hydrogen and muonic deuterium
and isotope shift $\Delta E_{IS}$.}
\bigskip
\begin{ruledtabular}
\begin{tabular}{|c|c|c|c|c|c|c|}  \hline
The contribution &\multicolumn{2}{|c|}{$\mu p$, meV}&
\multicolumn{2}{|c|}{$\mu d$, meV}&$\Delta E_{IS}$  & Ref. \\  \cline{2-3}\cline{4-5}
 & $1S$ & $2S$ & $1S$& $2S$ &  &  \\
to the energy of the atom& & & & & & \\  \hline
1&2&3&4&5&6&7\\  \hline
Fine structure& & & & & & \\
formula& & & & & & \\
$E_n=m_1+m_2-\frac{\mu^2(Z\alpha)^2}{2n^2}-$ &$1043927826$ &
$1043929722$&$1981268455$ &$1981270453$ &101030.3530 &(3)  \\
$-\frac{\mu(Z\alpha)^4}{2n^3} \left[1-\frac{3}{4n}+\frac{\mu^2}
{4nm_1m_2}\right]$ &$\sim$470.3586 &$\sim$866.0601 &$\sim$762.7537 &$\sim$188.8081 & &\cite{EGS}  \\   \hline
One-loop VP & & & & & & (12) \\
in $1\gamma$ interaction & & & & & &(13) \\
of order $\alpha(Z\alpha)^2$ &-1898.8379 &-219.5849 &-2129.2820 &-245.3205 &204.7085 &(14)  \\  \hline
Wichmann-Kroll& 0.0114 &0.0012 & 0.0126 & 0.0014 &-0.0010 & (16) \\
correction of & & & & & & (17) \\
order $\alpha(Z\alpha)^4$   & & & & & & \cite{WK} \\ \hline
Two-loop VP (VP-VP)  & & & & & & \\
in $1 \gamma$ interaction  & & & & & &(19)  \\
of order $\alpha^2(Z\alpha)^2$ &-1.8816 &-0.2426 &-2.1871 &-0.2811 &0.2616 &(20) \\  \hline
Two-loop VP (2-loop)  & & & & & & \\
in $1 \gamma$ interaction  & & & & & &(23)  \\
of order $\alpha^2(Z\alpha)^2$ &-12.6144 &-1.4112 &-14.0141 &-1.5606 &1.2476 &(24) \\  \hline
Three-loop VP (VP-  & & & & & & \cite{KN1} \\
VP-VP) in $1\gamma$ interaction  & & & & & & (27) \\
of order $\alpha^3(Z\alpha)^2$ &-0.0029 &-0.0003 &-0.0034 &-0.0004 &0.0004 &(28) \\  \hline
Three-loop VP (VP-  & & & & & & \cite{KN1} \\
-2-loop) in $1\gamma$ interaction  & & & & & &(29)  \\
of order $\alpha^3(Z\alpha)^2$ &-0.0223 &-0.0030 &-0.0251 &-0.0035 &0.0023 &(30) \\  \hline
Three-loop VP ($\Pi^{(p6)}$) & & & & & & \\
in $1\gamma$ interaction  & & & & & &  \\
of order $\alpha^3(Z\alpha)^2$ &-0.0340 &-0.0045 &-0.0380 &-0.0050 &0.0035 &\cite{KN1} \\  \hline
Relativistic and  & & & & & & \\
VP correction in  & & & & & & \cite{KP} \\
FOPT of order $\alpha^3(Z\alpha)^2$ &0.1962 &0.0249 &0.2515 &0.0322 &-0.0480 &(33) \\  \hline
Relativistic and VP  & & & & & & \\
effects in SOPT  & & & & & &\cite{KP}  \\
of order $\alpha^3(Z\alpha)^2$ &-0.2644 &-0.0559 &-0.3194 &-0.0696 &0.0413 &(43) \\  \hline
Two-loop VP in  & & & & & & (45)\\
SOPT of order $\alpha^2(Z\alpha)^2$ &-2.0343 &-0.1532 &-2.3675 &-0.1750 &0.3114 &(46) \\  \hline
Three-loop VP & & & & & & \\
(VP-VP,VP) in SOPT  & & & & & &(47)  \\
of order $\alpha^3(Z\alpha)^2$ &-0.0061 &-0.0002 &-0.0073 &-0.0005 &0.0009 &(49) \\  \hline
\end{tabular}
\end{ruledtabular}
\end{table}
\begin{table}[!ht]
Table I (continued).\\
\bigskip
\begin{ruledtabular}
\begin{tabular}{|c|c|c|c|c|c|c|}  \hline
1&2&3&4&5&6&7\\  \hline
Three-loop VP & & & & & & \\
(2-loop~VP,VP) in  & & & & & &(50)  \\
SOPT of order $\alpha^3(Z\alpha)^2$ &-0.0059 &-0.0016 &-0.0069 &-0.0021 &0.0005 &(51) \\  \hline
Muon SE and VP  & & & & & &(53) \\
of order $\alpha(Z\alpha)^4$ &5.1180 &0.6543 &5.9395 &0.7594 &-0.7164 &(54) \\  \hline
Radiative  & & & & & &(56) \\
corrections  & & & & & &(57) \\
of order $\alpha(Z\alpha)^5$ &0.0355 &0.0044 &0.0414 &0.0052 &-0.0051 &\cite{EGS} \\  \hline
Radiative and & & & & & &(58) \\
VP corrections  & & & & & &(59) \\
of order $\alpha^2(Z\alpha)^4$ &0.0178 &0.0025 &0.0209 &0.0029 &-0.0027 &(63),\cite{KP} \\  \hline
Recoil correction  & & & & & &(64) \\
of order $(Z\alpha)^5$ &0.3009 &0.0428 &0.1781 &0.0253 &0.1053 &\cite{EGS} \\  \hline
Nuclear structure & 38.5711 &4.8214 & 213.4218 & 26.6825 &-152.6597 & (66) \\
correction of order $(Z\alpha)^4$ & & & & & & \cite{EGS,FM3} \\  \hline
Nuclear structure &-0.1464&-0.0183&-2.9384&
-0.3674&2.4429& (69)  \\
correction of order $(Z\alpha)^5$ & & & & & & \cite{KP,FM3} \\  \hline
Nuclear structure &  &  &   &  &  &(70) \\
and VP correction &  &  &   &  &  & \\
of order $\alpha(Z\alpha)^4$ &0.2127 &0.0274 & 1.4155&0.1824&-1.0478&(71)\\  \hline
Nuclear structure & & & & & & \\
and VP correction in& & & & & &(72) \\
SOPT of order $\alpha(Z\alpha)^4$ &0.1327&0.0135&0.8913&0.0898&-0.6823&(73)  \\  \hline
Nuclear polarizability & & & & & &(74) \\
correction & & & & & &(75) \\
of order $(Z\alpha)^5$ &-0.1291&-0.0161&92.0511&11.5064&-80.6577 & \cite{MPK}  \\
 & & & & & &\cite{Friar,FM4} \\   \hline
HVP correction & & & & & &\cite{FMS} \\
of order $\alpha(Z\alpha)^4$ &-0.0864&-0.0108&-0.1010&-0.0126&0.0128 & \cite{FM2}  \\
 & & & & & & (77) \\  \hline
Summary &$1043927824$ &$1043929722$ &$1981268453$ &$1981270452$ & &   \\
contribution &$\sim$598.8893 &$\sim$650.1499 &$\sim$925.6873 &$\sim$980.2974 &101~003.3495 &   \\   \hline
\end{tabular}
\end{ruledtabular}
\end{table}

In this paper we calculate different quantum electrodynamic corrections,
effects of the nucleus structure and polarizability, hadronic vacuum polarization
to the Lamb shift of the $1S$ and $2S$ energy levels in muonic hydrogen,
muonic deuterium and isotope shift $(\mu p)$ - $(\mu d)$ for the splitting
$(1S-2S)$. The main goal of the work is to summarize various corrections
to the fine structure interval $(1S-2S)$ in the $(\mu p)$, $(\mu d)$ and
the muonic hydrogen - muonic deuterium isotope shift for this splitting
which would allow to obtain reliable theoretical values for the indicated
quantities. We consider the corrections of orders $\alpha^3$, $\alpha^4$,
$\alpha^5$ and also several important contributions of order $\alpha^6$
enhanced by the $\ln\alpha$. In our calculation we took into account that
the ratio $\mu\alpha/m_e$ is close to one and centred special attention
on the effects of the electron vacuum polarization. Numerical values of
obtained contributions are presented in Table I. We included also in Table I
the references on many papers where analytical or numerical calculations
of some corrections are performed despite the fact that particular numerical
results for the $1S$ and $2S$-levels are absent in these papers. For the
comparison of obtained results with the calculations carried out by other
authors we used commonly the review article \cite{EGS} which accumulate
recent advances in the physics of the energy spectra of simple atomic
systems and contain detailed references on the earlier performed investigations.

Let us point out some peculiarities of performed calculations:

1. The effects of the vacuum polarization for the considered muonic atoms
have played significant role. They lead to the modification of the Breit
two-particle interaction operator and give the corrections in the energy
spectra up to fifth order in $\alpha$.

2. The nuclear structure effects in the energy spectrum of the $S$-states
are expressed in terms of the nucleus (the proton and deuteron) charge
radius in the leading order $(Z\alpha)^4$ and in the next to leading order
$(Z\alpha)^5$ for the one-loop amplitudes by means of the nucleus
electromagnetic form factors.

3. The estimation of the nuclear polarizability contributions is performed
on the basis of Eq.(74) obtained by the Mil'shtein, Petrosyan and Khriplovich
and Eq.(75) in Refs.\cite{FM,KP1999}. The nuclear structure and polarizability
contributions lead to largest theoretical uncertainty for the transition
$(1S-2S)$ and the isotope shift $(\mu p)$ - $(\mu d)$.

Total numerical values for the energy levels $1S$ and $2S$ in muonic hydrogen
and muonic deuterium presented in Table I and also the values of the fine
structure interval $(1S-2S)$, isotope shift $(\mu p)$ - $(\mu d)$ can be
considered as a proper estimation for the future experiments with these
muonic atoms. The values of the corrections are obtained with the accuracy
0.0001 meV. For the splitting $(1S-2S)$ the error of theoretical result
is determined by a number of the factors. The uncertainties of fundamental
parameters (fine structure constant, the proton, deuteron and muon masses)
amount the value of order $10^{-7}$. The quantum electrodynamic corrections
of higher order $\alpha^6$ give theoretical error of order $10^{-8}$. The
largest contribution to the theoretical error is connected with the uncertainties
of the proton and deuteron charge radii. Their relative contributions reaches
the value of order $10^{-6}$ (we used the values of the proton and deuteron
charge radii: $r_p$ = 0.891 fm, $r_d$ = 2.094 fm). Further improvements
of theoretical results presented in Table I are related primarily with the
nuclear structure and polarizability corrections. So, for the comparison
of performed calculations with future experimental data a major
interest is connected with the quantity of fine structure interval
$[E(1S)-8E(2S)]$ which has not contain the nuclear structure and polarizability
corrections of the leading order $(Z\alpha)^4$. Numerical values for this
fine structure interval in muonic hydrogen and muonic deuterium are the
following:
\begin{equation}
\mu p:~~~E(1S)-8E(2S)=-7~307~509~956~602.3099~meV,
\end{equation}
\begin{displaymath}
\mu d:~~~E(1S)-8E(2S)=-13~868~895~169~916.6917~meV.
\end{displaymath}
The relative value of that part of theoretical error in Eq.(79) which is
determined by the QED corrections of higher order is extremely small (of
order $10^{-15}$).

The muonic hydrogen - muonic deuterium isotope shift for the splitting
$(1S-2S)$ is considered to be among the most important characteristics
of the energy spectra of muonic atoms. The differences in wavelengths of lines
emitted by isotopes of the same element can arise either as a result of the
differences in the masses of the isotopes or on account of differences in
the nuclear charge distributions. Whereas the masses of the proton and
deuteron are determined at present with sufficiently high accuracy, the
nuclear structure parameters are known less precisely. Using the results
of fulfilled calculations we can express the difference of the deuteron and
proton charge radii in terms of isotope shift:
\begin{equation}
\frac{r_d^2}{\left(1+\frac{m_\mu}{m_d}\right)^3}-\frac{r_p^2}
{\left(1+\frac{m_\mu}{m_p}\right)^3}=\frac{12}{7m_\mu^3(Z\alpha)^4}
\left(\Delta \tilde E^{th}_{IS}-\Delta E^{exp}_{IS}\right),
\end{equation}
where theoretical value $\Delta \tilde E^{th}_{IS}$ doesn't contain the
nuclear structure correction of order $(Z\alpha)^4$. So, the measurement
of the isotope shift $(\mu p)-(\mu d)$ will allow on the one hand to perform
additional test of the QED and on the other hand to obtain more accurate
value of the deuteron charge radius from the relation (80) after the
extraction the proton charge radius in the experiment at Paul Scherrer
Institute (PSI) \cite{K1}.

\begin{acknowledgments}
The author is grateful to D.D.Bakalov, M.I.Eides, R.N.Faustov,
V.V.Fil'chenkov, T.Kinoshita, I.B.Khriplovich, V.A.Matveev, M.Nio,
K.Pachucki, V.G.Pal'chikov, V.A.Rubakov, V.I.Savrin for useful discussions of
various aspects of the problem of muonic hydrogen energy spectrum. The work
was performed under the financial support of the Russian Fond for Basic
Researches (grant No. 04-02-16085).
\end{acknowledgments}

\end{document}